\documentclass[amsmath,amssymb,jap,superscriptaddress,twocolumn]{revtex4-1}

\usepackage{graphicx}
\usepackage{dcolumn}
\usepackage{bm}

\usepackage{float}
\usepackage{caption}
\usepackage{subcaption}

\begin{document}

\title{Revisiting Effects of Nitrogen Incorporation and Graphitization on Conductivity of Ultra-nano-crystalline Diamond Films}

\author{Tanvi Nikhar}
\email{nikharta@msu.edu}
\affiliation{Department of Electrical and Computer Engineering, Michigan State University, 428 S. Shaw Ln., East
Lansing, MI 48824, USA}
\author{Robert Rechenberg}
\affiliation{Fraunhofer USA Inc., Center for Coatings and Diamond Technologies
1449 Engineering Research Ct., B100
East Lansing, MI 48823, USA}
\author{Michael Becker}
\affiliation{Fraunhofer USA Inc., Center for Coatings and Diamond Technologies
1449 Engineering Research Ct., B100
East Lansing, MI 48823, USA}
\author{Sergey V. Baryshev}
\email{serbar@msu.edu}
\affiliation{Department of Electrical and Computer Engineering, Michigan State University, 428 S. Shaw Ln., East
Lansing, MI 48824, USA}

\begin{abstract}

Detailed structural and electrical properties of ultra-nano-crystalline diamond (UNCD) films grown in H$_\text{2}$/CH$_\text{4}$/N$_\text{2}$ plasma were systematically studied as a function of deposition temperature ($T_d$) and nitrogen content (\% N$_2$) to thoroughly evaluate their effects on conductivity. $T_d$ was scanned from 1000 to 1300 K for N$_2$ fixed at 0, 5, 10 and 20 \%. It was found that even the films grown in the synthetic gas mixture with no nitrogen could be made as conductive as 1--10$^{-2}$ $\Omega$ cm with overall resistivity of all the films tuned over 4 orders of magnitude through varying growth parameters. On a set of 27 samples, Raman spectroscopy and scanning electron microscopy show a progressive and highly reproducible film material phase transformation, from ultra-nano-crystalline diamond to nano-crystalline graphite as deposition temperature increases. The rate of this transformation is heavily dependent on N$_2$ content. Addition of nitrogen greatly increases the amount of $sp^2$ bonded carbon in the films thus enhancing the physical connectivity in the GB network that have high electronic density of states. However, addition of nitrogen greatly slows down crystallization of $sp^2$ phase in the GBs. Therefore, proper balance between GB connectivity and crystallinity is the key in conductivity engineering of (N)UNCD.

\end{abstract}

\maketitle

\section{\label{sec:level1}Introduction}

Diamond electronics faces great challenges due to the lack of low activation energy dopants. To allow room temperature operation, non-conventional concepts e.g. surface transfer doping \cite{Landstrass1989,Prins2000,Sakaguchi1999} have been developed. $p$-type doping with boron \cite{Grot1991} aided with surface transfer doping has helped achieve high performance devices with extremely low activation energy \cite{Williams2001}. Still, to date, problems with $n$-type doping of diamond \cite{KOIZUMI20009,Kato2005} have hindered its development as a ubiquitous device material. Doping of bulk diamond with nitrogen (common donor impurity) produces a deep level 1.7 eV below the conduction band minimum. Other $n$-type dopants, such as sulfur and phosphorus, have been attempted with only limited success as the room-temperature conductivity was too low for applying these materials in conventional electronic device architectures. A shallow donor species is still sought for diamond.

Ultra-nano-crystalline diamond (UNCD) is another form of synthetic diamond which is shown to exhibit outstanding and tunable $n$-type conductivity, and excellent biocompatibility \cite{auciello_gurman_guglielmotti_olmedo_berra_saravia_2014,article,Bajaj2007,Xiao2006InVA}, thus enabling a range of applications toward high power and bio-medical devices. Typically, conductivity is tuned through adding N$_\text{2}$ or ammonia gas to the precursor gas mixture. Nitrogen modifies the morphology \cite{Birrell2002} and leads to orders of magnitude change in the electrical conductivity\cite{Bhattacharyya2001}. This physical picture is rooted in theoretical models that predict grain boundaries (GB) contained in UNCD induce specific $sp^2$ electronic defect states in the fundamental band gap of $sp^3$ diamond. When nitrogen is added, it preferentially incorporates into the GB and further increases density of defect states turning them into bands capable of transferring the charge. At the same time, a strong morphology change \cite{Birrell2002, Bhattacharyya2001} resulting in enhanced GB size increases connectivity in the touching GB forming conducting networks \cite{Zapol2006}.

Despite significant progress in terms of applications, some fundamental aspects related to nitrogen incorporation into UNCD remained unaddressed. For example, the conductivity was found to saturate after attaining a certain nitrogen content in the plasma, which was further attributed to possible nitrogen incorporation saturation in UNCD \cite{Bhattacharyya2001,Achatz2006}. The specific nitrogen content in the precursor plasma required to obtain the maximum saturated conductivity value has large uncertainty \cite{Birrell2003, Achatz2006}. All the studies with varying nitrogen content were reported at a constant deposition temperature ($T_d$), and nitrogen was claimed to play the key role in setting high semi-metallic conductivity. Later, Ikeda $et$ $al.$ found strong dependence of conductivity on $T_d$. Addition of nitrogen played modest role \cite{Ikeda2008}. Most recently, Alcantar-Pena $et$ $al.$ \cite{Dallas2016} achieved exceptional conductivity tunability, up to 100 S/cm, in undoped (0\% N$_\text{2}$ in plasma) UNCD by simply varying $T_d$. Since both $T_d$ (through fundamental diamond-to-graphite phase transition) and N$_2$ content (through volume inflation upon incorporation) can promote $sp^2$ phase formation, the relative strength of both processes and corresponding material transformations merit additional investigations. The studies conducted in Refs.\cite{Ikeda2008,Dallas2016} were either restricted to limited number of $T_d$ or N$_2$ values.

The present work is motivated from the described inconsistencies and lack of complete understanding of the factors responsible for conduction in (N)UNCD films. The objective of this work is therefore to systematically vary both N$_2$ content and $T_d$ to revisit and better assess the physical origins transforming UNCD material from high to low resistivity state.

\section{\label{sec:level1}Experiment}

UNCD films were grown on ultrasonically seeded intrinsic Si (100) substrates using microwave plasma chemical vapor deposition technique in a reactor operated at 2.45 GHz \cite{DiamondFilmsHandbook}. Four different series of samples were grown from a hydrogen rich H$_\text{2}$/N$_\text{2}$/CH$_\text{4}$ feed gas mixture with 0, 5, 10, and 20$\%$ of nitrogen (volume \%), which corresponded to flow rates of 0, 10, 20 and 40 standard cubic centimeters per min (sccm) of N$_\text{2}$. The CH$_\text{4}$ flow rate was kept constant at 10 sccm (5$\%$ by volume), while the flow rate of N$_\text{2}$ and H$_\text{2}$ was varied to maintain a total flow rate of 200 sccm. Different deposition temperatures were achieved ranging 1043 to 1295 K by varying the total pressure of the vacuum chamber (35 to 60 torr) and the input microwave power (2.5 to 3.0 kW). The substrate temperature was measured using an infrared pyrometer during a growth process of 60 minutes for each sample.

All samples were characterized using Horiba Raman spectrometer with 532 nm probing laser. High resolution scanning electron microscopy (SEM) was performed using a JEOL JSM 7500F to study the surface and bulk morphology. Four-point probe measurements were used to calculate the resistivity of the films using the thickness calculated from (1) the weight measurements done before and after the growth process using a Mettler Toledo XS105 balance, and (2) reflectance interferogram measurements were performed using a Shimadzu UV-vis 2600 cross validated using SEM cross-section imaging. Surface roughness measurements were taken using Dektak 6M profilometer.

\section{\label{sec:level1}Results and Discussion}

The four different UNCD sample sets, totalling 27 samples, that were synthesized at four different nitrogen concentrations 0, 5, 10 and 20$\%$ are further labeled SA, SB, SC and SD, respectively. The following tables (Table \ref{table:zero}, \ref{table:five}, \ref{table:ten} and \ref{table:twenty} for series SA, SB, SC and SD respectively) summarize the resultant properties of each series of films and the values of input parameters used to tune $T_d$. As seen, control of $T_d$ was achieved by varying the input parameters \emph{viz.} the forward microwave power and total pressure in the chamber. An increase in the substrate temperature was observed as nitrogen was increased in the plasma (replacing hydrogen) while all other input parameter values were kept fixed. Since the amount of nitrogen in the plasma dominated the substrate temperature, this effect could be explained in terms of how the energy is transferred from the gas molecules/ions to the substrate. N$_2$ is a heavier molecule compared to H$_2$, and thus the energy transfer resulting from the bombardment of the substrate with N$_2$ is more efficient leading to higher $T_d$.

\begin{widetext}

\begin{table}[H]
\resizebox{\textwidth}{!}{%
\begin{tabular}{|c|c|c|c|c|c|c|c|c|c|c|c|c|}
\hline
\textbf{Sample} & \textbf{\begin{tabular}[c]{@{}l@{}}t$_\text{weight}$ \\ (nm)\end{tabular}} & \textbf{\begin{tabular}[c]{@{}l@{}} t$_\text{UV-vis}$ \\ (nm)\end{tabular}} & \textbf{\begin{tabular}[c]{@{}l@{}}Sheet \\ Resistance \\ R$_\text{s}$ (k$\Omega$/$\square$)\end{tabular}} & \textbf{\begin{tabular}[c]{@{}l@{}}Resistivity \\ ($\Omega$cm)\end{tabular}} & \textbf{\begin{tabular}[c]{@{}l@{}}Avg. \\ Roughness\\R$_\text{a}$ (nm)\end{tabular}} & \textbf{\begin{tabular}[c]{@{}l@{}}Peak to peak \\ Roughness \\R$_\text{z}$ (nm)\end{tabular}} & \textbf{\begin{tabular}[c]{@{}l@{}}Deposition \\ Temperature \\ (Kelvin)\end{tabular}} & \textbf{\begin{tabular}[c]{@{}l@{}}Deposition \\ Time (hrs)\end{tabular}} & \textbf{\begin{tabular}[c]{@{}l@{}}Total \\ Pressure \\ (sccm)\end{tabular}}  & \textbf{CH$_4$} & \textbf{N$_2$} & \textbf{\begin{tabular}[c]{@{}l@{}}Forward \\ Power \\ (W)\end{tabular}} \\ \hline
SA1 & 463 & 499.6 & Too high      &        & 1.44 & 8.85  & 1048 & 1 & 40 & 5\% & 0\% & 2500 \\ \hline
SA2 & 659 & 550.4 & 3.5x10$^\text{3}$ & 192.64 & 2.82 & 20.85 & 1098 & 1 & 45 & 5\% & 0\% & 3000 \\ \hline
SA3 & 657 &  NA   & 30.8          & 2.025  & 1.66 & 9.22  & 1126 & 1 & 50 & 5\% & 0\% & 3000 \\ \hline
SA4 & 532 &  NA   & 6.4           & 0.341  & 1.5  & 8.14  & 1156 & 1 & 55 & 5\% & 0\% & 3000 \\ \hline
SA5 & 525 &  NA   & 0.297         & 0.015  & 2.11 & 10.61 & 1213 & 1 & 57 & 5\% & 0\% & 3000 \\ \hline
SA6 & 687 &  NA   & 0.128         & 0.009  & 3.19 & 18.02 & 1248 & 1 & 60 & 5\% & 0\% & 3000 \\ \hline
\end{tabular}%
}
\caption{Series SA films grown with 0$\%$ N$_\text{2}$ concentration}\label{table:zero}
\end{table}
\begin{table}[H]
\resizebox{\textwidth}{!}{%
\begin{tabular}{|c|c|c|c|c|c|c|c|c|c|c|c|c|}
\hline
\textbf{Sample} & \textbf{\begin{tabular}[c]{@{}l@{}}t$_\text{weight}$ (nm)\end{tabular}} & \textbf{\begin{tabular}[c]{@{}l@{}} t$_\text{UV-vis}$ (nm)\end{tabular}} & \textbf{\begin{tabular}[c]{@{}l@{}}Sheet \\ Resistance \\ R$_\text{s}$ (k$\Omega$/$\square$)\end{tabular}} & \textbf{\begin{tabular}[c]{@{}l@{}}Resistivity \\ ($\Omega$cm)\end{tabular}} & \textbf{\begin{tabular}[c]{@{}l@{}}Avg. \\ Roughness\\R$_\text{a}$ (nm)\end{tabular}} & \textbf{\begin{tabular}[c]{@{}l@{}}Peak to peak \\ Roughness \\R$_\text{z}$ (nm)\end{tabular}} & \textbf{\begin{tabular}[c]{@{}l@{}}Deposition \\ Temperature \\ (Kelvin)\end{tabular}} & \textbf{\begin{tabular}[c]{@{}l@{}}Deposition \\ Time (hrs)\end{tabular}} & \textbf{\begin{tabular}[c]{@{}l@{}}Total \\ Pressure \\ (sccm)\end{tabular}}  & \textbf{CH$_4$} & \textbf{N$_2$} & \textbf{\begin{tabular}[c]{@{}l@{}}Forward \\ Power \\ (W)\end{tabular}} \\ \hline
SB1 & 456 & 393.52 & 15x10$^\text{3}$     & 590.2  & 4.2  & 19.34 & 1043 & 1 & 35 & 5\% & 5\% & 2500 \\ \hline
SB2 & 541 & 528.9  & 5x10$^\text{3}$      & 264.4  & 4.47 & 24.61 & 1085 & 1 & 40 & 5\% & 5\% & 2500 \\ \hline
SB3 & 498 & 518.4  & 12.56      & 0.65   & 4.83 & 18.78 & 1103 & 1 & 40 & 5\% & 5\% & 2700 \\ \hline
SB4 & 636 & 584.37 & 20.63      & 1.2055 & 1.88 & 13.55 & 1118 & 1 & 45 & 5\% & 5\% & 2500 \\ \hline
SB5 & 595 & 539.5  & 20.5 & 1.106  & 1.7  & 12.02 & 1125 & 1 & 45 & 5\% & 5\% & 2700 \\ \hline
SB6 & 479    &   NA     & 3.76       & 0.1803 & 1.83 & 14.9  & 1148 & 1 & 43 & 5\% & 5\% & 3000 \\ \hline
SB7 & 481 &   NA     & 5.29       & 0.2544 & 4.51 & 21.6  & 1151 & 1 & 45 & 5\% & 5\% & 3000 \\ \hline
SB8 & 453 &   NA     & 2.74       & 0.124  & 2.09 & 9.93  & 1163 & 1 & 50 & 5\% & 5\% & 3000 \\ \hline
SB9 & 532 &   NA     & 0.17       & 0.009  & 2.76 & 14.79 & 1229 & 1 & 60 & 5\% & 5\% & 3000 \\ \hline
\end{tabular}%
}
\caption{Series SB films grown with 5$\%$ N$_\text{2}$ concentration}\label{table:five}
\end{table}
\begin{table}[H]
\resizebox{\textwidth}{!}{%
\begin{tabular}{|c|c|c|c|c|c|c|c|c|c|c|c|c|}
\hline
\textbf{Sample} & \textbf{\begin{tabular}[c]{@{}l@{}}t$_\text{weight}$ (nm)\end{tabular}} & \textbf{\begin{tabular}[c]{@{}l@{}} t$_\text{UV-vis}$ (nm)\end{tabular}} & \textbf{\begin{tabular}[c]{@{}l@{}}Sheet \\ Resistance \\ R$_\text{s}$ (k$\Omega$/$\square$)\end{tabular}} & \textbf{\begin{tabular}[c]{@{}l@{}}Resistivity \\ ($\Omega$cm)\end{tabular}} & \textbf{\begin{tabular}[c]{@{}l@{}}Avg. \\ Roughness\\R$_\text{a}$ (nm)\end{tabular}} & \textbf{\begin{tabular}[c]{@{}l@{}}Peak to peak \\ Roughness \\R$_\text{z}$ (nm)\end{tabular}} & \textbf{\begin{tabular}[c]{@{}l@{}}Deposition \\ Temperature \\ (Kelvin)\end{tabular}} & \textbf{\begin{tabular}[c]{@{}l@{}}Deposition \\ Time (hrs)\end{tabular}} & \textbf{\begin{tabular}[c]{@{}l@{}}Total \\ Pressure \\ (sccm)\end{tabular}}  & \textbf{CH$_4$} & \textbf{N$_2$} & \textbf{\begin{tabular}[c]{@{}l@{}}Forward \\ Power \\ (W)\end{tabular}} \\ \hline
SC1 & 391 & 344.69 & 4x10$^\text{3}$ & 137.87 & 2.13 & 19.02 & 1068 & 1 & 35 & 5\% & 10\% & 2500 \\ \hline
SC2 & 500 & 484.13 & 42.9  & 2.077  & 2.68 & 20.51 & 1101 & 1 & 40 & 5\% & 10\% & 2500 \\ \hline
SC3 & 627 & 595.27 & 50.38 & 2.999  & 2.27 & 22.7  & 1145 & 1 & 45 & 5\% & 10\% & 2700 \\ \hline
SC4 & 497 & 513.58 & 13.5  & 0.69   & 1.88 & 16.72 & 1163 & 1 & 45 & 5\% & 10\% & 3000 \\ \hline
SC5 & 517 &   NA   & 0.66  & 0.034  & 2.39 & 12.17 & 1223 & 1 & 55 & 5\% & 10\% & 3000 \\ \hline
\end{tabular}%
}
\caption{Series SC films grown with 10$\%$ N$_\text{2}$ concentration}\label{table:ten}
\end{table}
\begin{table}[H]
\resizebox{\textwidth}{!}{%
\begin{tabular}{|c|c|c|c|c|c|c|c|c|c|c|c|c|}
\hline
\textbf{Sample} & \textbf{\begin{tabular}[c]{@{}l@{}}t$_\text{weight}$ (nm)\end{tabular}} & \textbf{\begin{tabular}[c]{@{}l@{}} t$_\text{UV-vis}$ (nm)\end{tabular}} & \textbf{\begin{tabular}[c]{@{}l@{}}Sheet \\ Resistance \\ R$_\text{s}$ (k$\Omega$/$\square$)\end{tabular}} & \textbf{\begin{tabular}[c]{@{}l@{}}Resistivity \\ ($\Omega$cm)\end{tabular}} & \textbf{\begin{tabular}[c]{@{}l@{}}Avg. \\ Roughness\\R$_\text{a}$ (nm)\end{tabular}} & \textbf{\begin{tabular}[c]{@{}l@{}}Peak to peak \\ Roughness \\R$_\text{z}$ (nm)\end{tabular}} & \textbf{\begin{tabular}[c]{@{}l@{}}Deposition \\ Temperature \\ (Kelvin)\end{tabular}} & \textbf{\begin{tabular}[c]{@{}l@{}}Deposition \\ Time (hrs)\end{tabular}} & \textbf{\begin{tabular}[c]{@{}l@{}}Total \\ Pressure \\ (sccm)\end{tabular}}  & \textbf{CH$_4$} & \textbf{N$_2$} & \textbf{\begin{tabular}[c]{@{}l@{}}Forward \\ Power \\ (W)\end{tabular}} \\ \hline
SD1 & 194 & 223.16 & 30x10$^\text{3}$ & 669.48 & 2.44 & 21.51 & 1115 & 1 & 35 & 5\% & 20\% & 2500 \\ \hline
SD2 & 279 & 239.66 & 12x10$^\text{3}$ & 287.6  & 2.13 & 18.61 & 1138 & 1 & 40 & 5\% & 20\% & 2500 \\ \hline
SD3 & 371 & 325.31 & 2x10$^\text{3}$  & 65.06  & 2.81 & 21.01 & 1195 & 1 & 45 & 5\% & 20\% & 3000 \\ \hline
SD4 & 367 & 398.96 & 19.7   & 0.786  & 3.53 & 27.86 & 1220 & 1 & 50 & 5\% & 20\% & 3000 \\ \hline
SD5 & 552 &   NA   & 8.36   & 0.46   & 3.15 & 18.5  & 1259 & 1 & 55 & 5\% & 20\% & 3000 \\ \hline
SD6 & 479 &   NA   & 4.7    & 0.225  & 2.79 & 18.9  & 1263 & 1 & 60 & 5\% & 20\% & 3000 \\ \hline
SD7 & 365 &   NA   & 1.45   & 0.053  & 2.56 & 14.73 & 1295 & 1 & 65 & 5\% & 20\% & 3000 \\ \hline
\end{tabular}%
}
\caption{Series SD films grown with 20$\%$ N$_\text{2}$ concentration}\label{table:twenty}
\end{table}
\end{widetext}

\subsection{Resistivity}

\begin{figure}[t]
\centering
\includegraphics[width =0.48\textwidth]{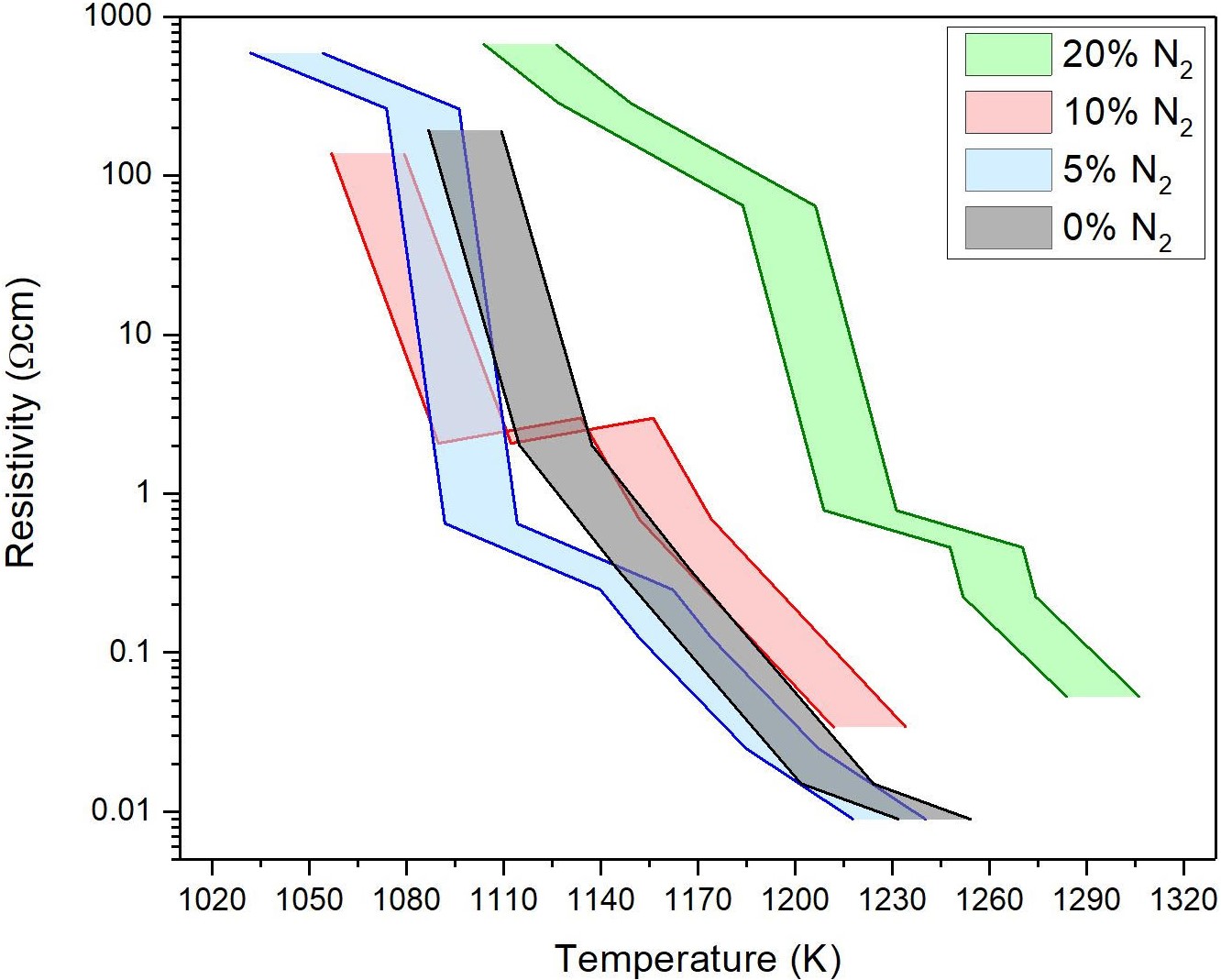}
\caption{Resistivity of all the films of series SA, SB, SC and SD as a function of deposition temperature } 
\label{fig:Resistivity}
\end{figure}

The room temperature resistivity of the films as a function of deposition temperature $T_d$ measured with an optical pyrometer is shown in Fig.~\ref{fig:Resistivity}. A variation of $\pm$10 K ($\Delta T_{d1}$) was observed based on different viewing angles along with a standard temperature variation of $\pm$5 K ($\Delta T_{d2}$) during the one hour growth. A resultant error bar of $\pm$11.2 K for the $x$--axis is thus calculated taking both these variations into account, as $(\Delta T_{d1}^2 + \Delta T_{d2}^2)^{1/2}$.

Fig.~\ref{fig:Resistivity} presents a direct comparison of resistivity variation for undoped and nitrogen incorporated UNCD films synthesized at different $T_d$. The resistivity decreases exponentially by more than four orders of magnitude from $\sim$10$^\text{2}$ to 10$^\text{-2}$ $\Omega$ cm as $T_d$ changes by about 25\%. The exponential decay hits the bottom at a resisitivity that corresponds to a conductivity of $\sim$100 S/cm. Room temperature (RT) conductivity of 100 S/cm is the champion conductivity value of (N)UNCD films claimed to be found only at N$_2$ concentration $>$10\% in the precursor plasma \cite{Bhattacharyya2001,Williams2004}. Important note is that the RT conductivity of (N)UNCD films has been shown insensitive to the ambient temperature \cite{Bhattacharyya2001, Zapol2006}, which is a characteristic semi-metallic behaviour. In our case, the champion conductivity can be achieved for any nitrogen concentration, even 0\%, through finding proper $T_d$. This result may look unusual but is supported by a few extensive experimental reports that widely varied the deposition temperature at constant N$_2$ concentrations, see for instance Refs.\cite{Dallas2016,Ikeda2008}. Ikeda $et$ $al.$ \cite{Ikeda2008} concluded that $sp^2$ phase in GB (dependent upon $T_d$) had to play significant role in setting conductivity. Pondering upon literature results \cite{Birrell2002, Birrell2003}, it becomes clear that N$_2$ incorporation affects conductivity by increasing the $sp^2$ content in GB. It is important to note that while the trend and resistivity values recorded in Fig.1 are in accordance with the increasing $T_d$ for a constant N$_2$ percentage, it is not the same for the increasing N$_2$ content for a constant $T_d$. It can be seen from Fig.~\ref{fig:Resistivity} that for a given $T_d$, the higher the N$_2$ the higher the resistivity -- this contradicts previous findings \cite{Bhattacharyya2001,Williams2004}.

Since it is the GBs that were found to be a common effect and source of inconsistencies including the newly found, it was hypothesized that the actual physical structure of GBs driven by thermo-chemical processes during plasma growth (rather than global plasma parameters such as $T_d$  and N$_2$) must be related to the resistivity. Therefore, a closer look is taken at Raman spectroscopy response of UNCD to evaluate diamond $sp^3$ and especially GB $sp^2$ transformations caused by the change of $T_d$ and N$_2$.

\subsection{Raman Spectroscopy}

Indeed, Raman spectroscopy is the most prominent method in determining structure of carbon-based electronic materials. The Raman of (N)UNCD features two major optical bands called D and G. Monitoring I(D)/I(G) ratio was proposed for diamond-to-graphite ratio evaluation [refs1,2,3].

Other studies [refs1,2,3] provided large sets of Raman spectra that showed remarkable material transformations in that the intensities of the D and G bands non-monotonically changed as N$_2$ and $T_d$ was varied. It revealed that the (N)UNCD material is highly "dynamic" upon being synthesized. Careful analysis of literature material [refs1,2,3] yields that not only intensities of the D and G bands change but so do their spectral line positions. For example, there is a clear and steady shift of the G band peak towards higher wavenumbers as $T_d$ increases. To the best of our knowledge this effect has never been systematically studied and addressed.

The Raman spectra collected for all the four series SA, SB, SC and SD exhibit some prominent common features described as follows

\noindent 1) A band with the peak position ranging between 1333 cm$^{-1}$ and 1350 cm$^{-1}$. The 1333 cm$^{-1}$ peak is the characteristic diamond peak attributed to $sp^3$ bonded carbon. The 1350 cm$^{-1}$ feature appears for graphitic materials in the form of a band induced by the defect or disorder in the $sp^2$ bonded carbon. Polycrystalline graphite exhibits D band in its Raman spectrum due to the outer rim of graphite crystallites which tends to have more defects as opposed to the inside of the graphite crystal for which the band is not present \cite{StudyingDisorder}. In this discussion, this band will be referred to as the D band as has been used traditionally for systems containing mixed phases of $sp^2$ and $sp^3$ bonded carbon: D band contains two D's in its name, one is for $sp^3$ (D)iamond and another one is for $sp^2$ (D)efect. In further support of this statement, the films of the SD series allow for resolving two features in the D band as can be seen in Fig.\ref{fig:Raman_N2_20}: one feature centers at 1333 cm$^{-1}$ and is interpreted to be due to $sp^3$ bonded carbon and the second one centers at 1364 cm$^{-1}$ and is due to disordered $sp^2$ bonded carbon.

\noindent 2) A band with the peak position ranging between 1545 cm$^\text{-1}$ and 1590 cm$^\text{-1}$ which can be recognized as the G band. In addition to the D band, graphitic materials also exhibit a band at 1588 cm$^\text{-1}$ called the G band attributed to the crystalline graphitic structure with $sp^2$ bonded carbon. The integrated intensity ratio of the D and G bands, I(D)/I(G), is directly proportional to the defect quantity in graphitic $sp^2$ materials and inversely proportional to the in-plane crystallite size.

\noindent 3) A shoulder at 1140 cm$^{-1}$ and a weak peak at 1480 cm$^{-1}$ representing $\omega_1$ and $\omega_3$ modes of trans-polyacetylene (t-PA) chains respectively.

The Raman spectra studied both as a function of $T_d$ and N$_2$ percentage show various trends in the position, intensity and shape of the D, G, $\omega_1$ and $\omega_3$ peaks discussed as follows.

\begin{figure}
\centering
\includegraphics[width =0.48\textwidth]{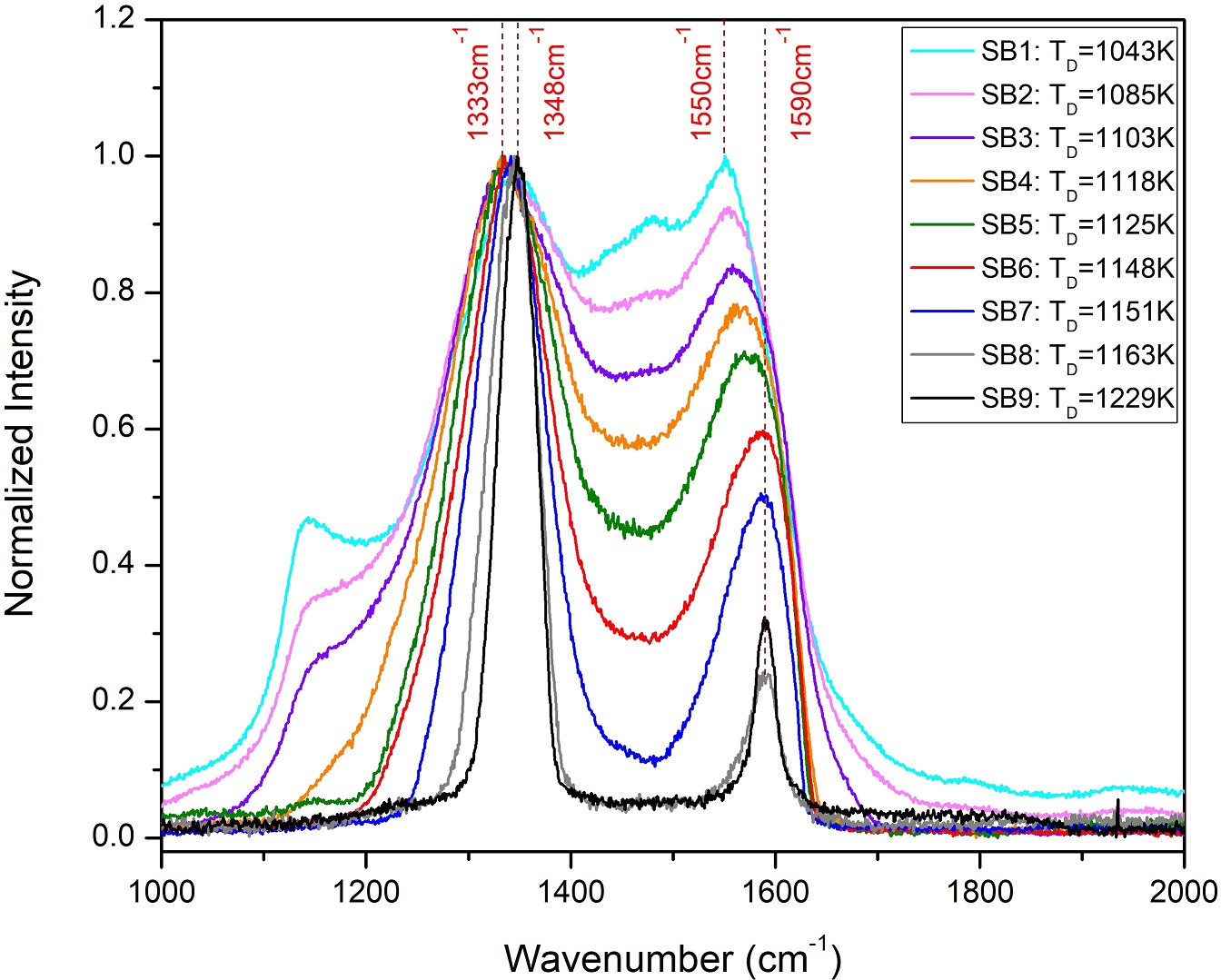}
\caption{Raman spectra for 5$\%$ N$_\text{2}$ samples as a function of T$_\text{d}$} 
\label{fig:Raman_N2_5}
\end{figure}

\begin{figure}
\centering
\includegraphics[width =0.48\textwidth]{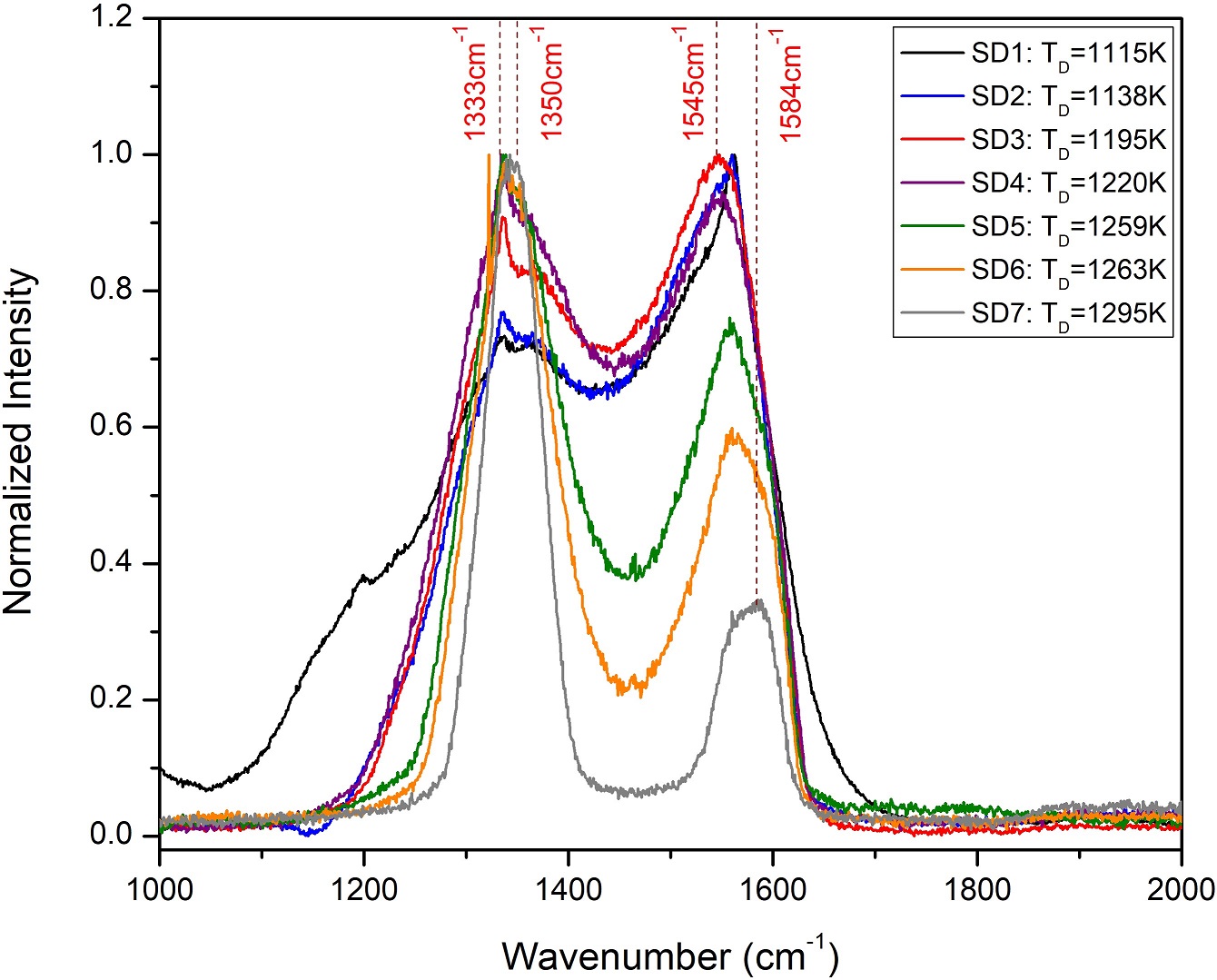}
\caption{Raman spectra for 20$\%$ N$_\text{2}$ samples as a function of T$_\text{d}$} 
\label{fig:Raman_N2_20}
\end{figure}

\subsubsection{Raman spectra as a function of $T_d$}

Fig.~\ref{fig:Raman_N2_5} shows the Raman spectra of films from SB series, grown with 5\% N$_2$ in the feed gas mixture, for various $T_d$. As the temperature increases, the G band peak shifts from 1550 cm$^{-1}$ to 1590 cm$^{-1}$ with decreasing intensity which shows a reduction of the amorphous carbon content and its conversion from short-range to long range order \cite{Ferrari_Robertson2000}. The peak shift is also accompanied by its narrowing indicating an increase in the crystallinity of the material. Thus, with the increase in $T_d$, the $sp^2$ bonded carbon changes from amorphous to polycrystalline graphite. The high intensity of defect-induced D band with respect to G band can be attributed to a large volume of edges resulting from small size GBs lining the diamond grains. 

For lower $T_d$, the D band is broad with its peak weighted at 1333 cm$^{-1}$. As $T_d$ increases to up to 1148 K, the band gets narrower while its peak spectral position stays constant at 1333 cm$^{-1}$. Beyond 1148 K, the D band continues to get narrower and the peak at 1333 cm$^{-1}$ instantaneously moves to 1350 cm$^\text{-1}$, indicating that there is no $sp^3$ carbon to be detected by Raman spectroscopy. The most striking transformation takes place within the SD series (Fig.~\ref{fig:Raman_N2_20}). The 1333 cm$^{-1}$ diamond peak keeps intensifying with respect to the 1350 cm$^{-1}$ peak as $T_d$ increases upto 1195 K which shows the most optimal condition to synthesize (N)UNCD films of the largest fraction of diamond content. For $T_d >$1195 K, the peaks merge together and shift completely to 1350 cm$^{-1}$ at 1295 K. This complete D band peak shift from 1333 cm$^{-1}$ to 1350 cm$^{-1}$ taking place at higher synthesis temperatures suggests the UNCD material converts from diamond $sp^3$ host matrix containing graphitic GBs to crystallized nano-graphite -- this is normal to expect due to the existence of fundamental phase transition at $\sim$1500 K from metastable diamond structure to its fundamental ground state which is graphite. The fact that this phase transformation happens at lower temperature is due to ultra-small ($\sim$10 nm) diamond grains enhancing the kinetic rate. The phenomenon of UNCD converting into other carbon structures when exposed to high temperatures, is also in agreement with previous results \cite{Dallas2016,Field_emission_properties_of_nanocrystalline_chemically_vapor_deposited_diamond_films}.

The spectral features at 1140 cm$^{-1}$ and 1480 cm$^{-1}$ de-intensify as $T_d$ increases and completely disappear after 1103 K: these are $\omega_1$ (in-plane C-H bending combined with C-C stretching) and the $\omega_3$ (C=C stretching) vibration modes of t-PA chains. Accordingly, these features were associated with fragments of t-PA-like species also present in the grain boundaries. The disappearance of these features due to t-PA decomposition also confirms that these modes of t-PA are not related to diamond grains \cite{Pfeiffer2003}. The removal of t-PA chains present in the grain boundaries also aids to the formation of better $sp^2$ bonded phase in GBs.

\begin{figure}[t]
\centering
\includegraphics[width =0.48\textwidth]{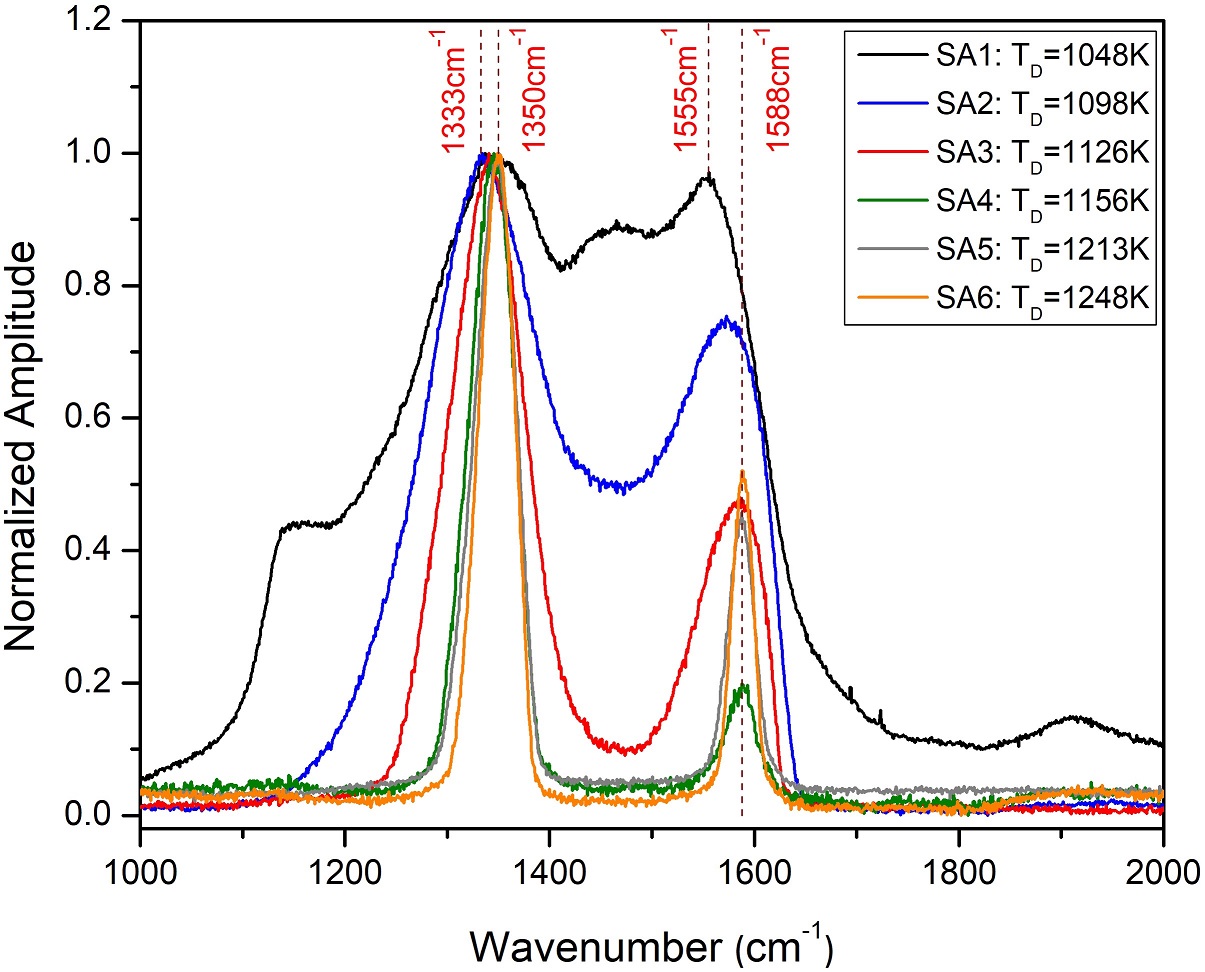}
\caption{Raman spectra for 0$\%$ N$_\text{2}$ samples as a function of T$_\text{d}$} 
\label{fig:Raman_N2_0}
\end{figure}

\begin{figure}[t]
\centering
\includegraphics[width =0.48\textwidth]{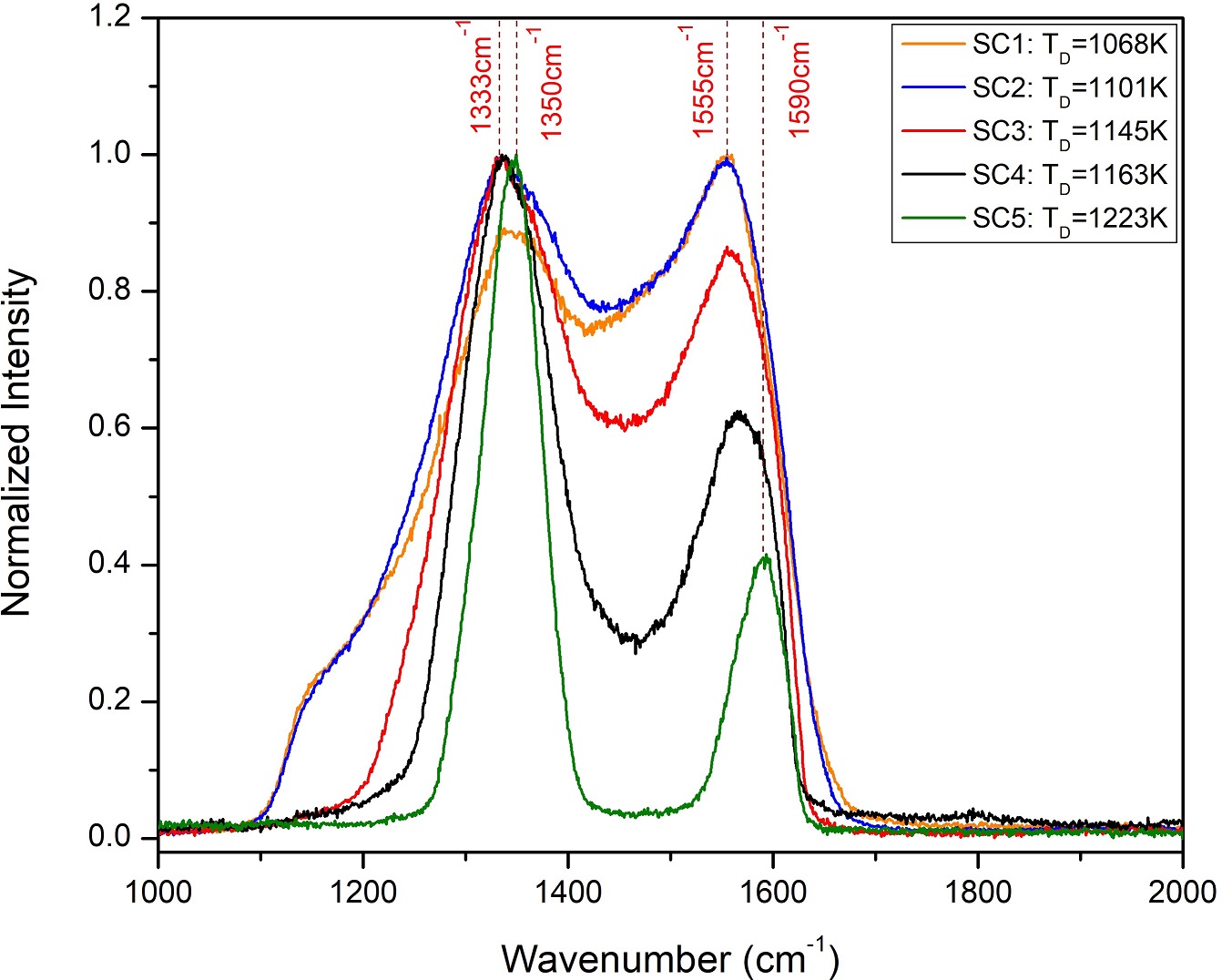}
\caption{Raman spectra for 10$\%$ N$_\text{2}$ samples as a function of T$_\text{d}$} 
\label{fig:Raman_N2_10}
\end{figure}

Quenching of t-PA spectral features was observed for all other sets, SA, SC and SD, to happen at near 1100 K, and was a convenient benchmark reference for temperature reliability measurement. Likewise, as seen in Fig.~\ref{fig:Raman_N2_0}, \ref{fig:Raman_N2_10} and \ref{fig:Raman_N2_20}, the trends of the D and G bands of Raman spectra of other three sets were consistent with the results outlined in regards to Fig.~\ref{fig:Raman_N2_5}. For increasing $T_d$ irrespective of the concentrations of nitrogen in the plasma, the common trends, yielding consistency behind the argument on material transformation from UNCD to nano-crystalline graphite, can be summarized as follows

\noindent 1) The shift in the D band peak from 1333 cm$^{-1}$ to 1350 cm$^{-1}$ along with the narrowing of the band.

\noindent 2) The decrease in the intensity and the shift in the G band peak from 1550 cm$^{-1}$ to 1590 cm$^{-1}$ along with the narrowing of the band.

\noindent 3) The decrease and quench of the t-PA features at 1140 cm$^{-1}$ and 1480 cm$^{-1}$. These features disappear after about 1100 K for all the four series of films which again confirms that these are not related to diamond.

\subsubsection{Raman spectra as a function of N$_2$ in the feed gas}

\begin{figure}
\centering
\includegraphics[width =0.48\textwidth]{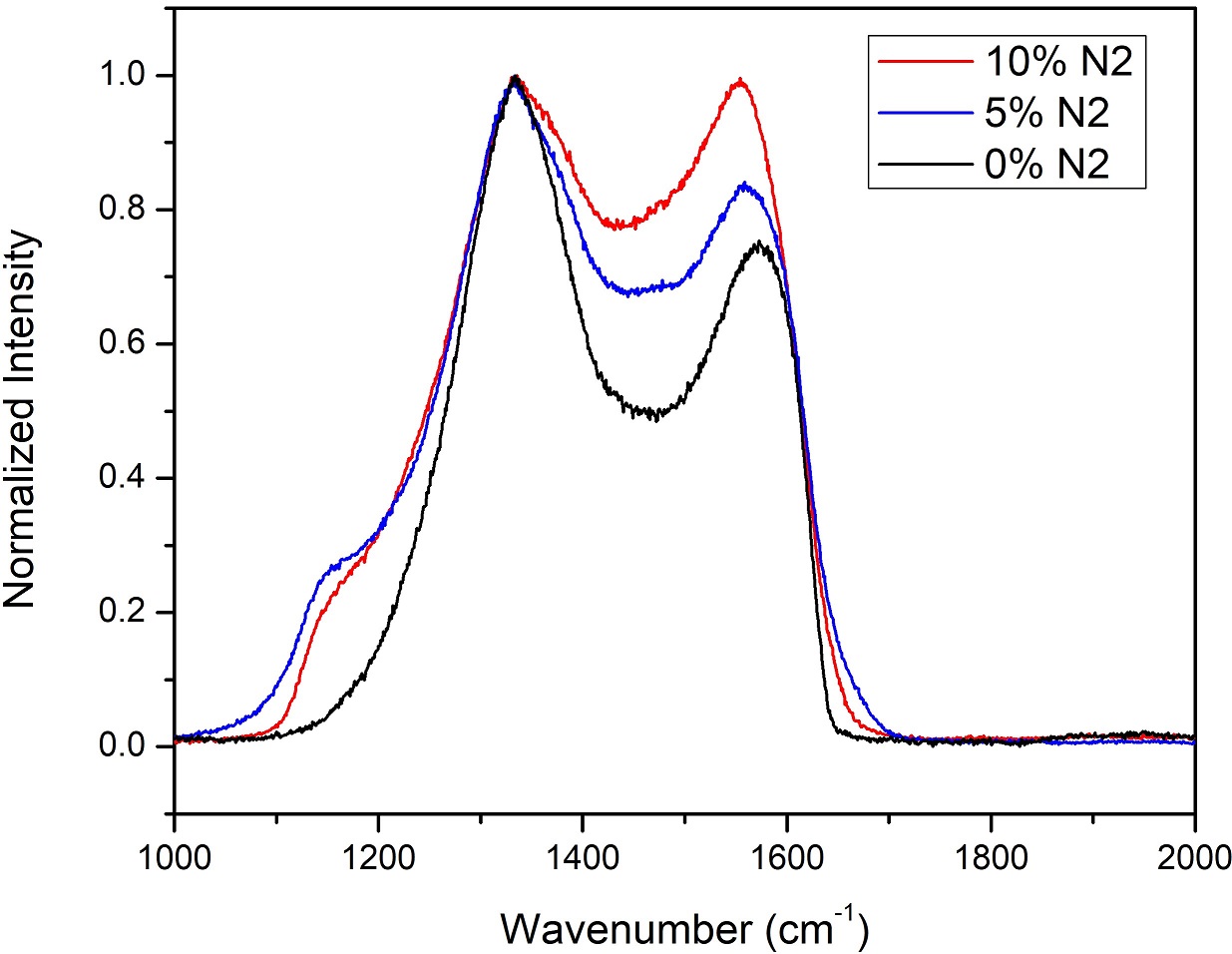}
\caption{Raman spectra for samples deposited at 1098 K with 0, 5 and 10$\%$ N$_\text{2}$ concentration} 
\label{fig:1098K}
\end{figure}

\begin{figure}
\centering
\includegraphics[width =0.48\textwidth]{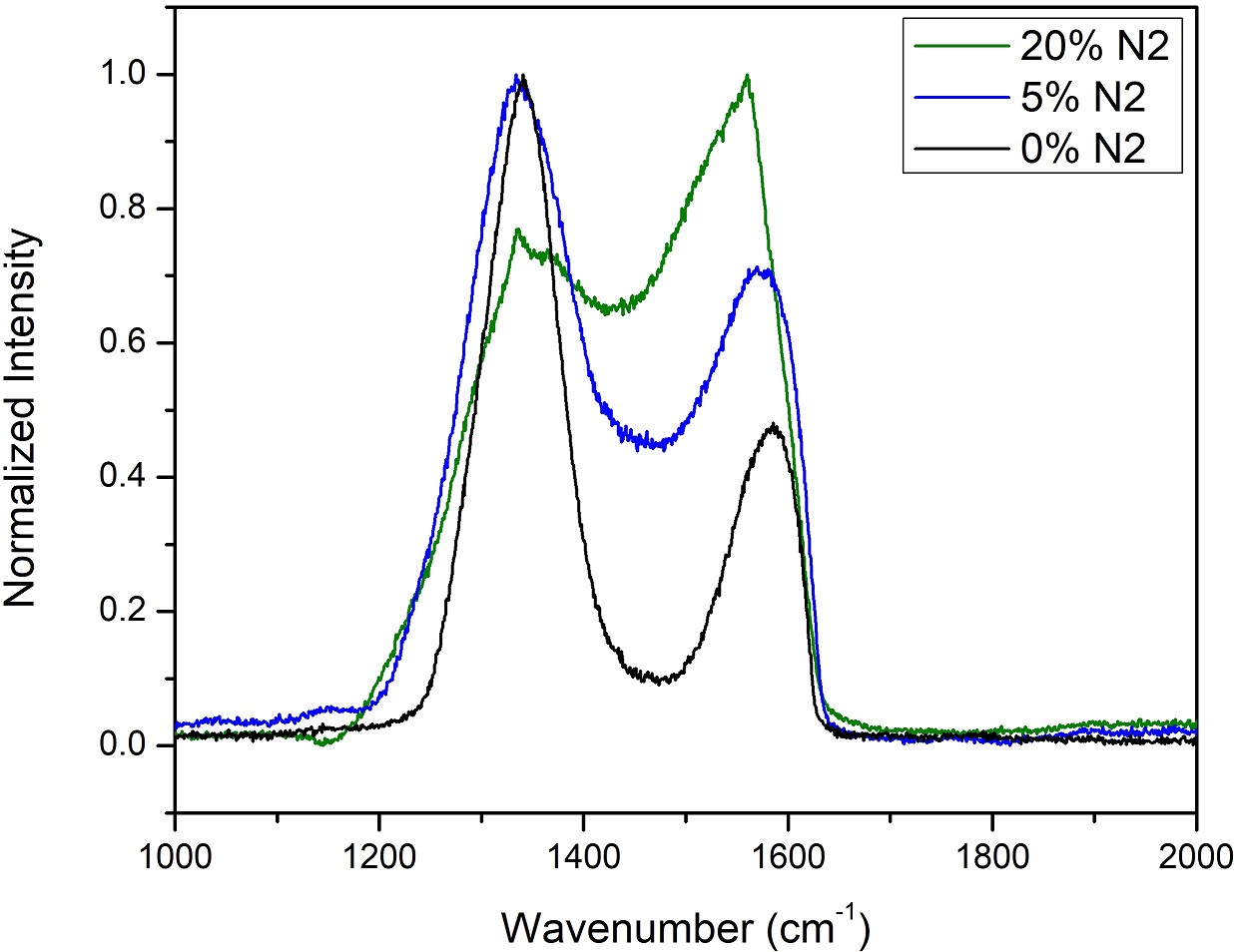}
\caption{Raman spectra for samples deposited at 1126 K with 0, 5 and 20$\%$ N$_\text{2}$ concentration} 
\label{fig:1126K}
\end{figure}

Figures \ref{fig:1098K} and \ref{fig:1126K} demonstrate comparison sets of Raman spectra for (N)UNCD films that were grown at the same $T_d$ of 1098 K and 1126 K respectively, but with varied N$_2$ content. For $T_d$ kept constant, the comparison of the peak intensity of the G band shows that the amount of $sp^2$ phase increases with increasing N$_2$ concentration in the growth plasma. Combined with the result in Fig.~\ref{fig:Raman_N2_20}, it does show that both $sp^2$ GB size and $sp^3$ diamond grain size are inflated by adding more nitrogen \cite{Birrell2002}. There is an important $nuance$ that exists: as the G peak increases with N$_2$ content, its position moves toward smaller wavenumbers. It means that even though the relative size of GB (i.e. the available conduction channel) is increased, such GBs are not of good use as they are largely amorphous. Therefore, it is concluded that N$_2$ alone is not a prerequisite for engineering high conductivity in 10\% and 20\% films. This is consistent with the major result presented in Fig.~\ref{fig:Resistivity}.

GB amorphization induced by the increasing nitrogen content can be thought of as result of the changing amount of energy required in the formation of increasing amount of CN bonds. The positive enthalpy of formation of many CN bonds \cite{CNEnthalpy2015} could make the incorporation of nitrogen in the GBs of UNCD an endothermic process. Thus, the available thermal energy in the CVD reactor needs to be increased via increasing $T_d$ to enable (1) the incorporation of nitrogen through the formation of CN bonds in the GB of the films, and (2) $sp^2$ GB crystallization. In other words, increasing CN bond formation for the incorporation of nitrogen leads to an overall increase in the required thermal energy to achieve the same level of crystallinity of sp$^\text{2}$ bonded carbon which is the major contributing factor towards the conductivity of the films. This hypothesis could account for (1) the increased amount of amorphous $sp^2$ phase in high N$_2$ \% films and (2) the upward shift in $T_d$ required to obtain (N)UNCD films with similar resistivities for increasing nitrogen content in the plasma. For instance, the resistivity value of $\sim$0.6 $\Omega\cdot$cm was achieved at 1103 K with 5$\%$ N$_\text{2}$ as opposed to 1163 K for 10$\%$ N$_\text{2}$ and 1221 K for 20$\%$ N$_\text{2}$.

\subsection{Scanning electron microscopy}
\begin{figure*}[htbp]
\centering
    \begin{subfigure}{0.3\textwidth}
        \includegraphics[width=\linewidth]{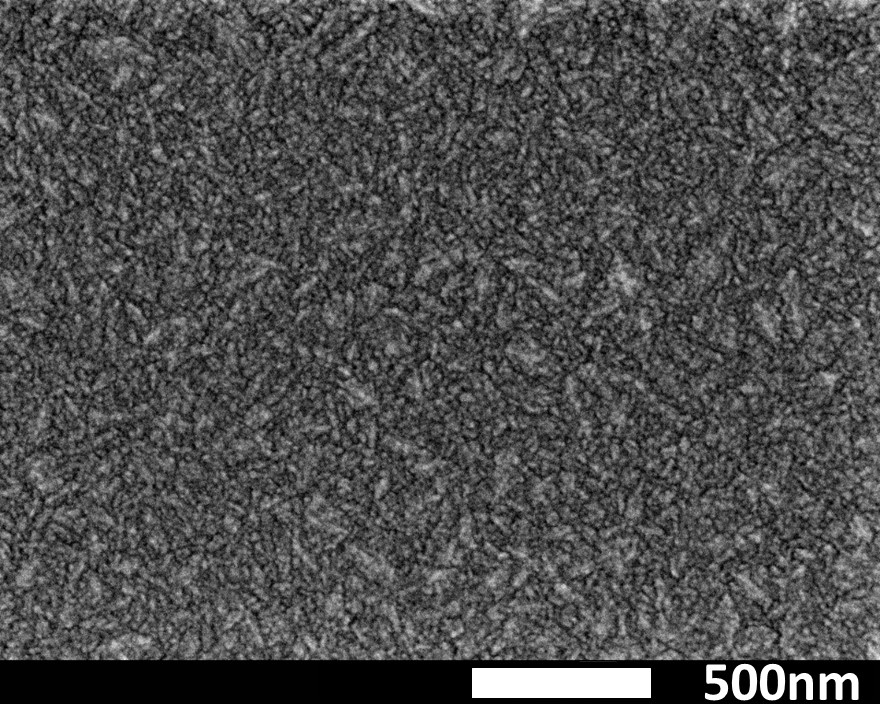}
         \caption{SA1: T$_\text{d}$=1048 K}
    \end{subfigure}
    \begin{subfigure}{0.3\textwidth}
        \includegraphics[width=\linewidth]{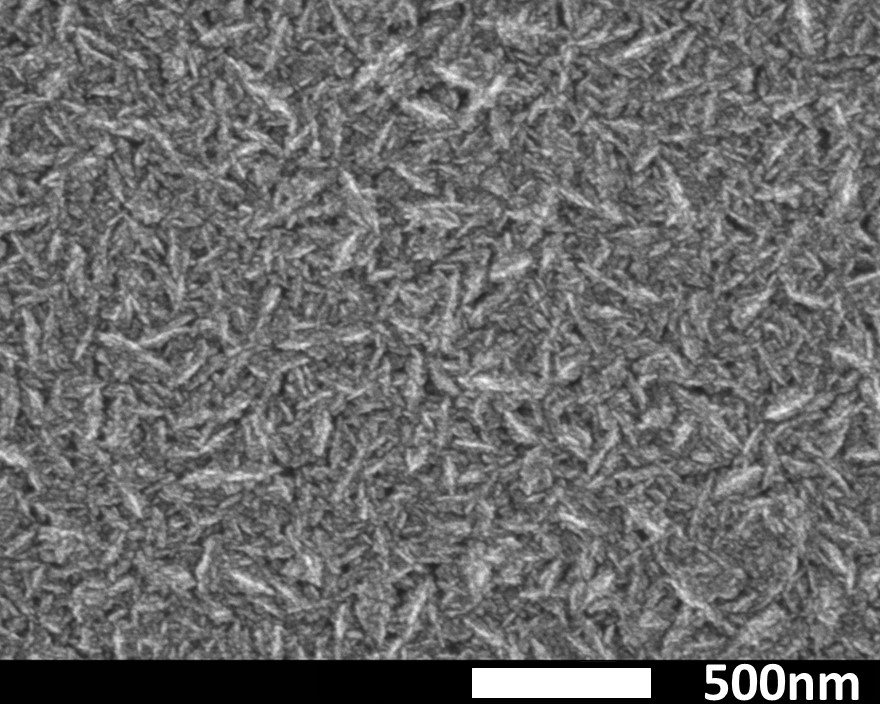}
        \caption{SA3: T$_\text{d}$=1126 K} 
    \end{subfigure}
    \begin{subfigure}{0.3\textwidth}
        \includegraphics[width=\linewidth]{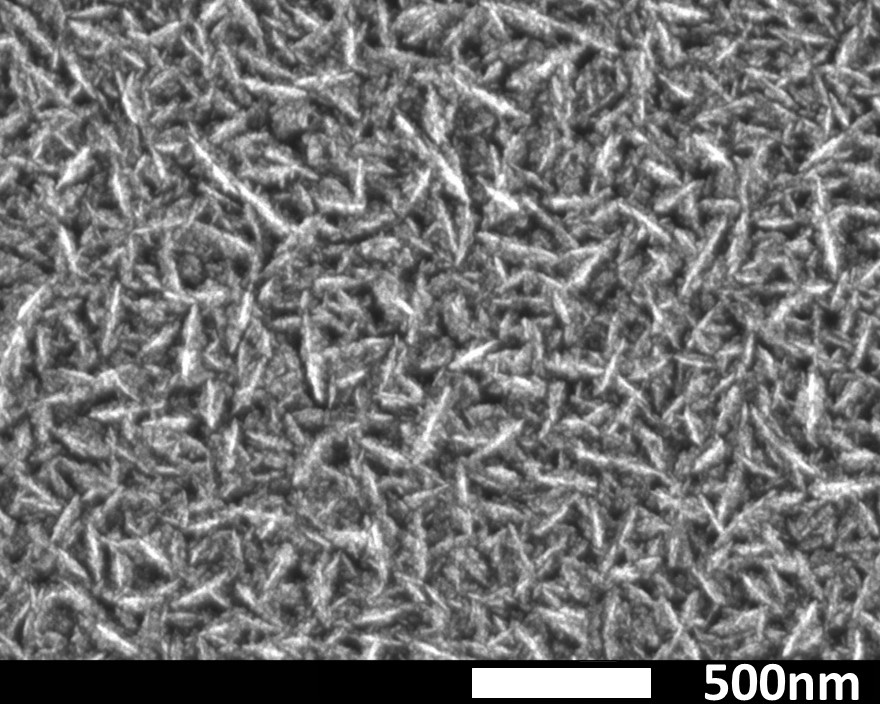}
        \caption{SA4: T$_\text{d}$=1156 K}
   \end{subfigure}
    \vspace{-0pt}
\caption{SEM images of samples grown with 0$\%$ N$_\text{2}$ concentration}
\label{fig:SEM_N2_0}
\vspace{0pt}

    \begin{subfigure}{0.3\textwidth}
        \includegraphics[width=\linewidth]{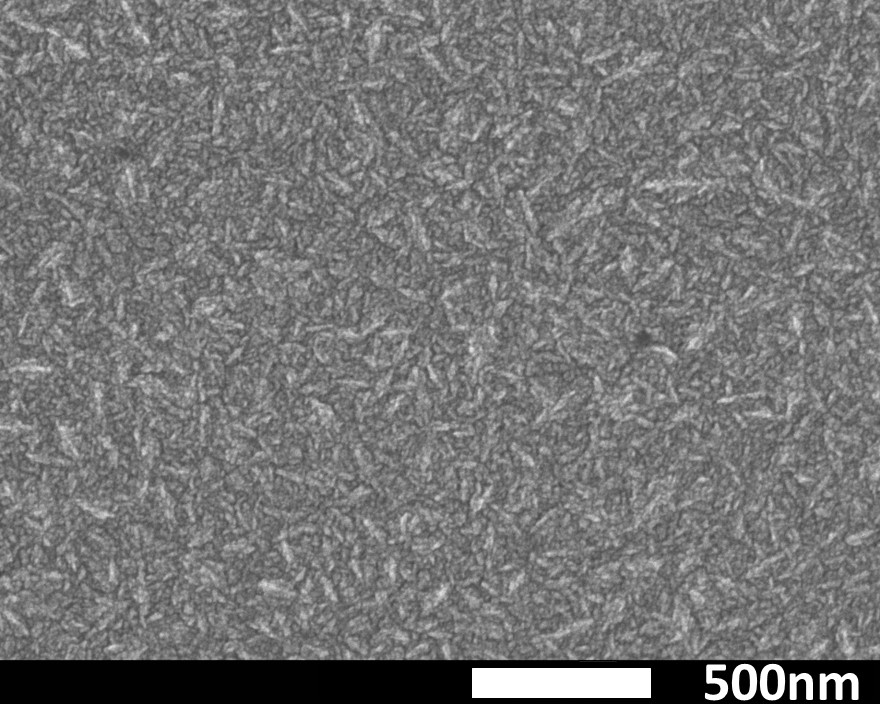}
        \caption{SB1: T$_\text{d}$=1043 K}
    \end{subfigure}
    \begin{subfigure}{0.3\textwidth}
        \includegraphics[width=\linewidth]{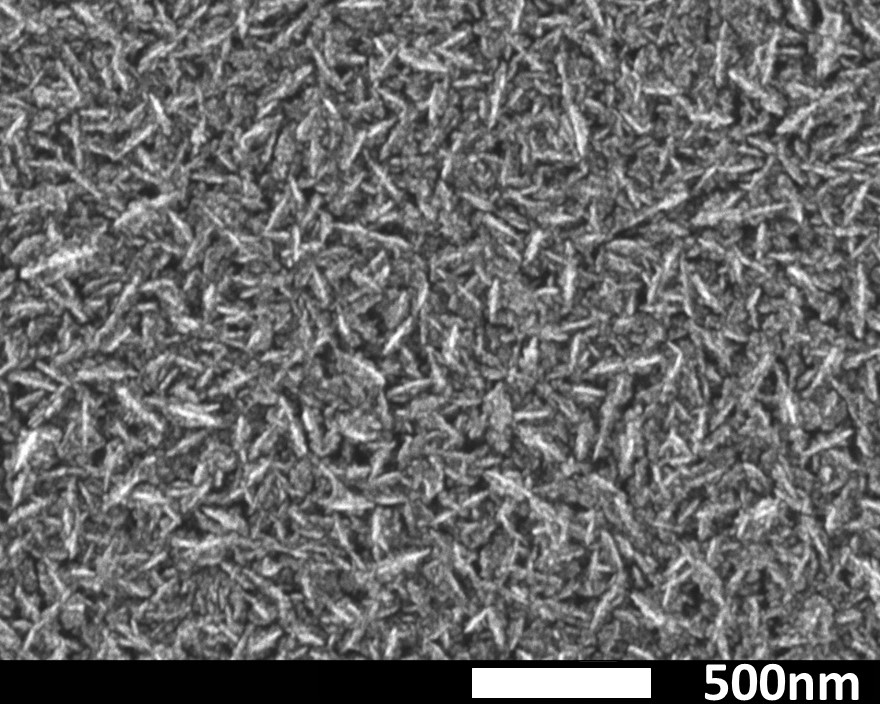}
        \caption{SB7: T$_\text{d}$=1151 K}
    \end{subfigure}
    \begin{subfigure}{0.3\textwidth}
        \includegraphics[width=\linewidth]{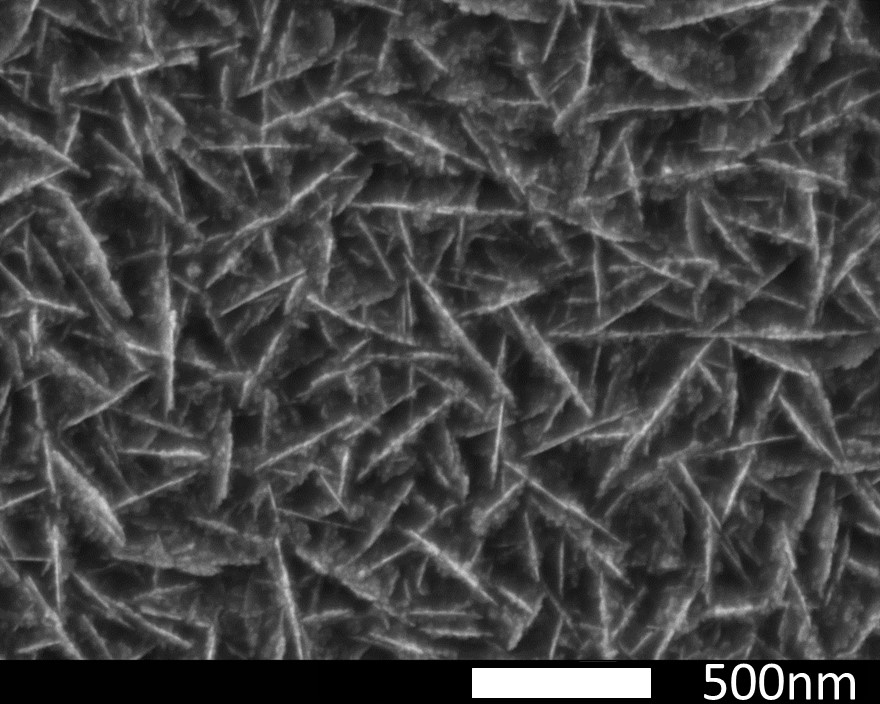} 
        \caption{SB9: T$_\text{d}$=1229 K}
   \end{subfigure}
    \vspace{-0pt}
\caption{SEM images of samples grown with 5$\%$ N$_\text{2}$ concentration}
\label{fig:SEM_N2_5}
\vspace{0pt}

    \begin{subfigure}{0.3\textwidth}
        \includegraphics[width=\linewidth]{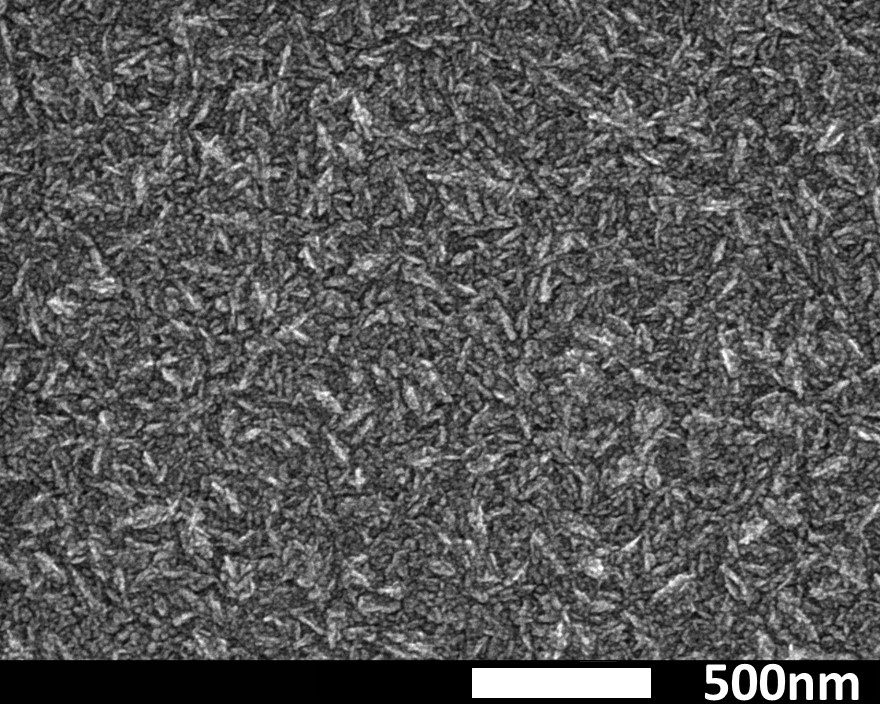} 
        \caption{SC2: T$_\text{d}$=1101 K}
    \end{subfigure}
    \begin{subfigure}{0.3\textwidth}
        \includegraphics[width=\linewidth]{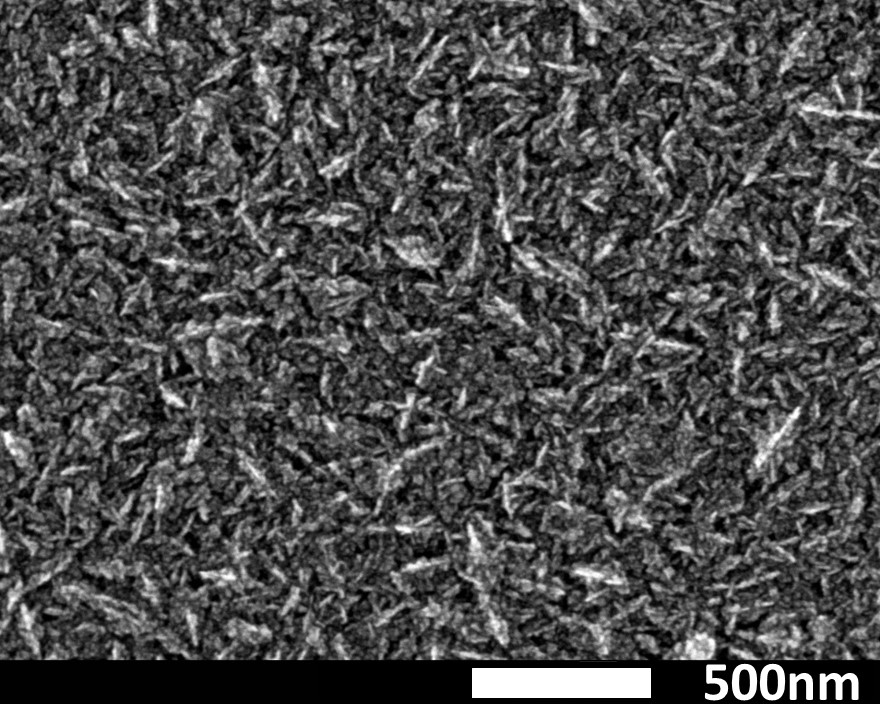}
        \caption{SC4: T$_\text{d}$=1163 K}
    \end{subfigure}
    \begin{subfigure}{0.3\textwidth}
        \includegraphics[width=\linewidth]{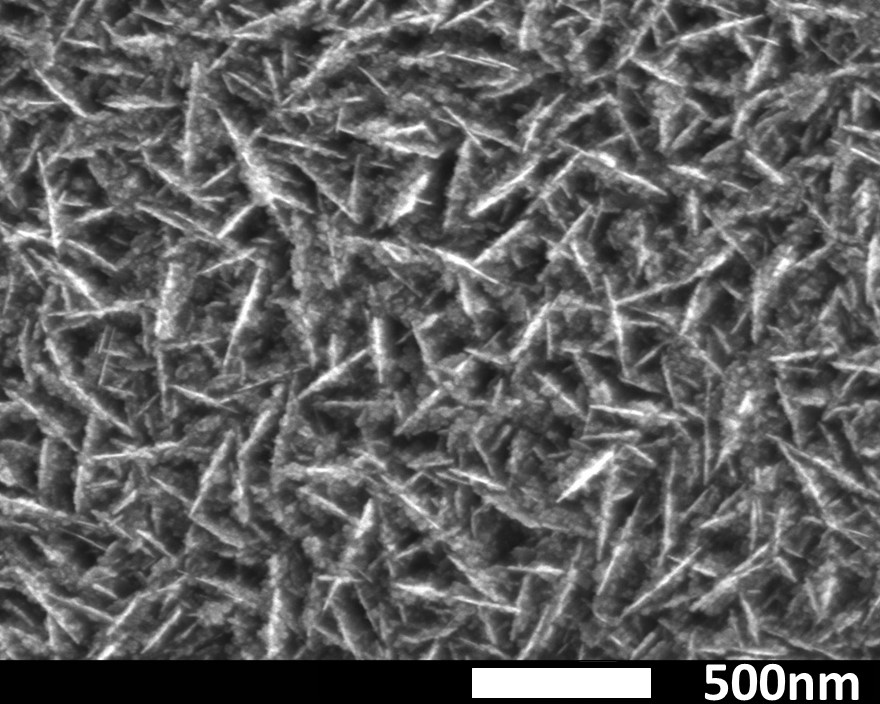} 
        \caption{SC5: T$_\text{d}$=1223 K}
    \end{subfigure}
    \vspace{-0pt}
\caption{SEM images of samples grown with 10$\%$ N$_\text{2}$ concentration}
\label{fig:SEM_N2_10}
\vspace{0pt}

    \begin{subfigure}{0.3\textwidth}
        \includegraphics[width=\linewidth]{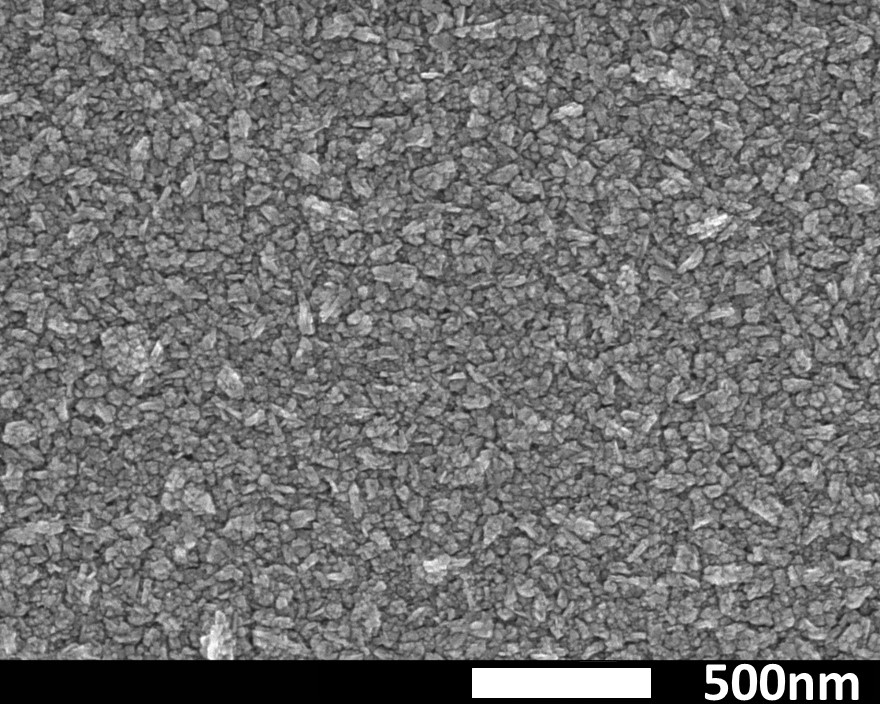}
        \caption{SD1: T$_\text{d}$=1115 K}
    \end{subfigure}   
    \begin{subfigure}{0.3\textwidth}
        \includegraphics[width=\linewidth]{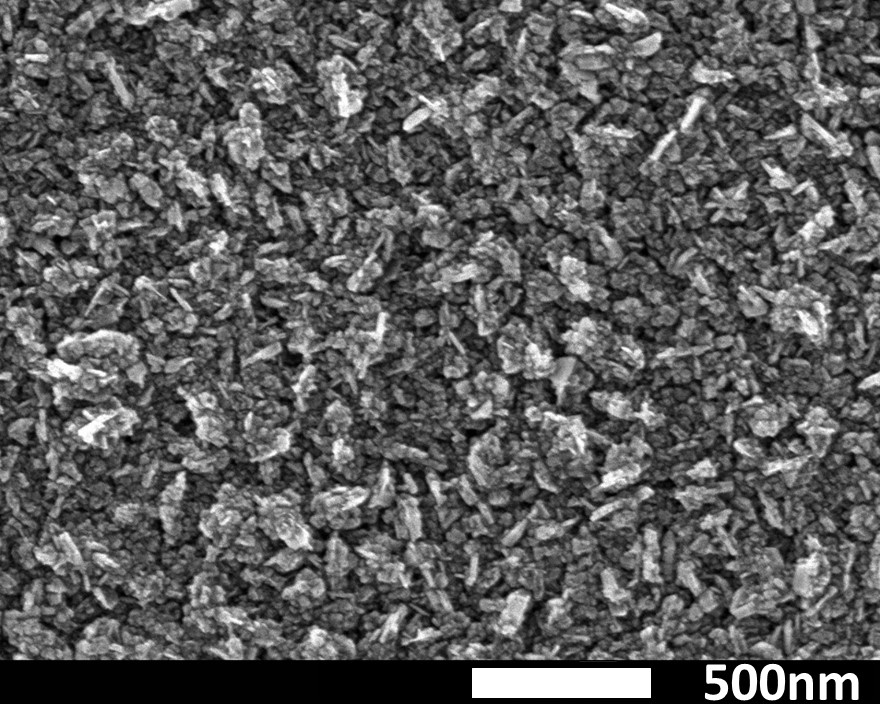}
        \caption{SD4: T$_\text{d}$=1220 k}
    \end{subfigure}   
    \begin{subfigure}{0.3\textwidth}
        \includegraphics[width=\linewidth]{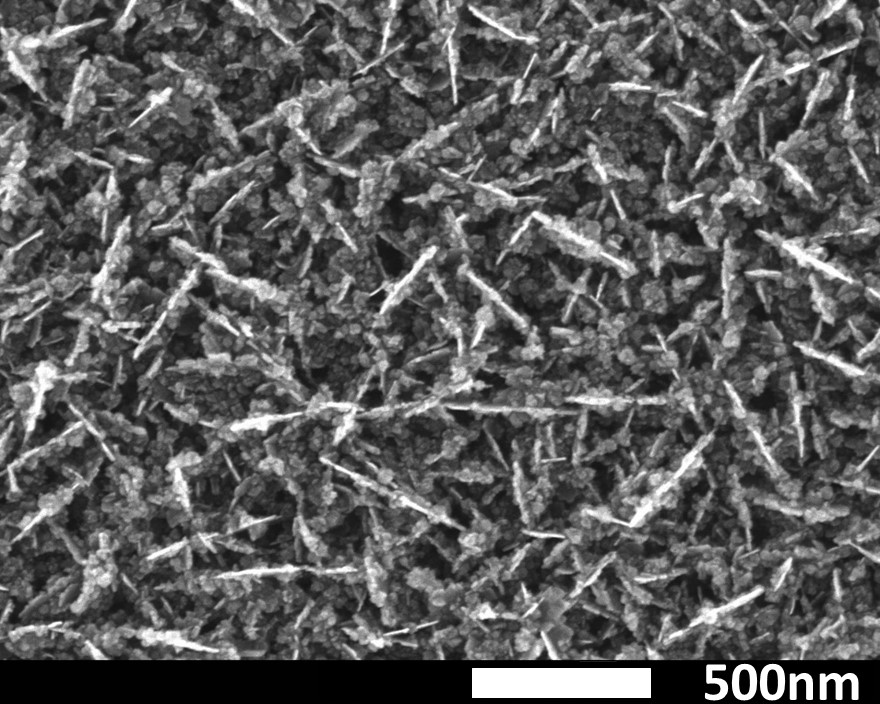} 
        \caption{SD7: T$_\text{d}$=1295 K}
     \end{subfigure}
     \vspace{-0pt}
\caption{SEM images of samples grown with 20$\%$ N$_\text{2}$ concentration}
\label{fig:SEM_N2_20}
\vspace{0pt}
\end{figure*}

Extensive data set of electron micrographs of the surface of the films further corroborates the main conclusions and observations that came from Raman spectroscopy and add more insight into details of (N)UNCD growth. Figures \ref{fig:SEM_N2_0}, \ref{fig:SEM_N2_5}, \ref{fig:SEM_N2_10} and \ref{fig:SEM_N2_20} illustrate systematic and reproducible surface morphology of films that modifies with increasing $T_d$ for 0, 5, 10 and 20$\%$ N$_2$ respectively.

\begin{figure*}[htbp]
\centering
    \begin{subfigure}{0.3\textwidth}
        \includegraphics[width=\linewidth]{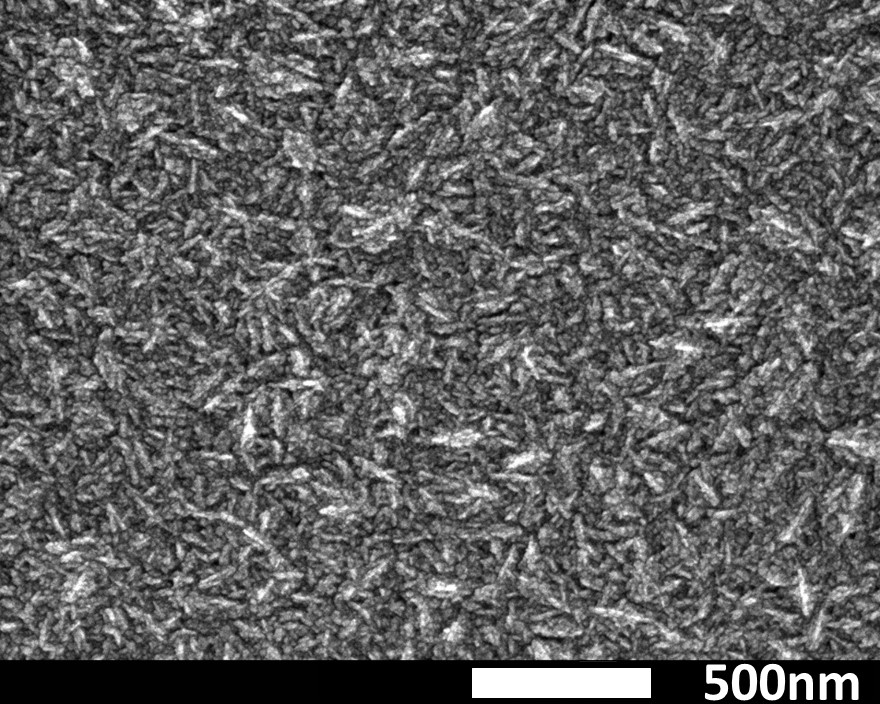} 
        \caption{Top view}
    \end{subfigure}
    \begin{subfigure}{0.3\textwidth}
        \includegraphics[width=\linewidth]{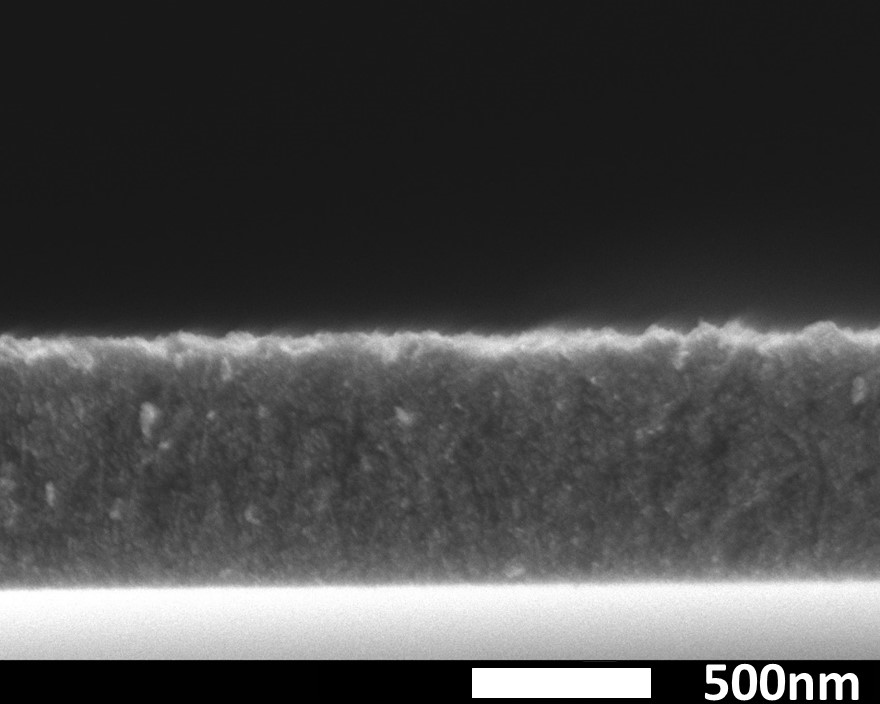}
        \caption{Cross-section}
    \end{subfigure}
    \begin{subfigure}{0.3\textwidth}
        \includegraphics[width=\linewidth]{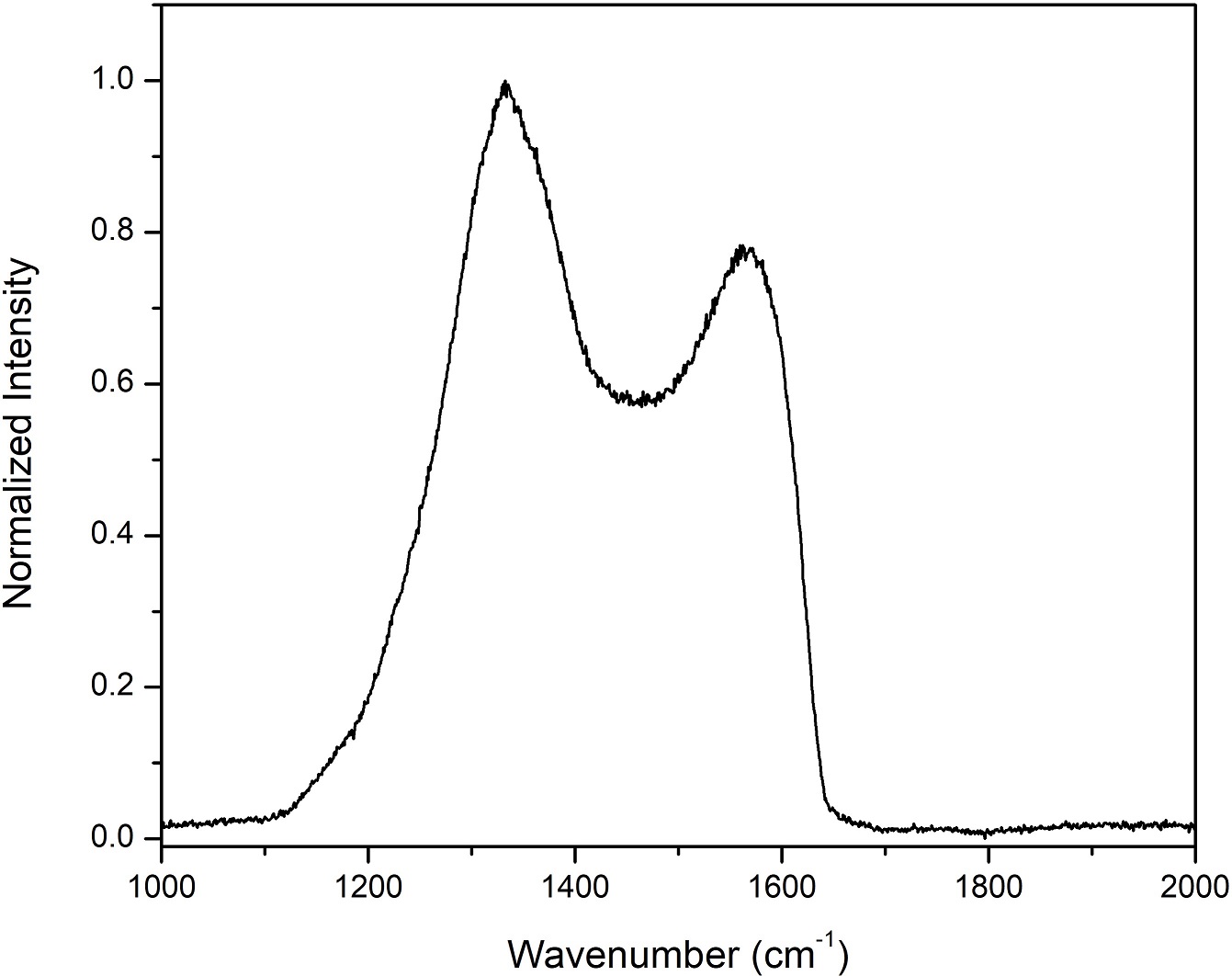}
        \caption{Raman spectrum}
    \end{subfigure}
    \vspace{-0pt}
\caption{SEM and Raman spectrum of sample:~SB4, grown with 5$\%$ N$_\text{2}$}
\label{fig:SB4}
\vspace{-0pt}

    \begin{subfigure}{0.3\textwidth}
        \includegraphics[width=\linewidth]{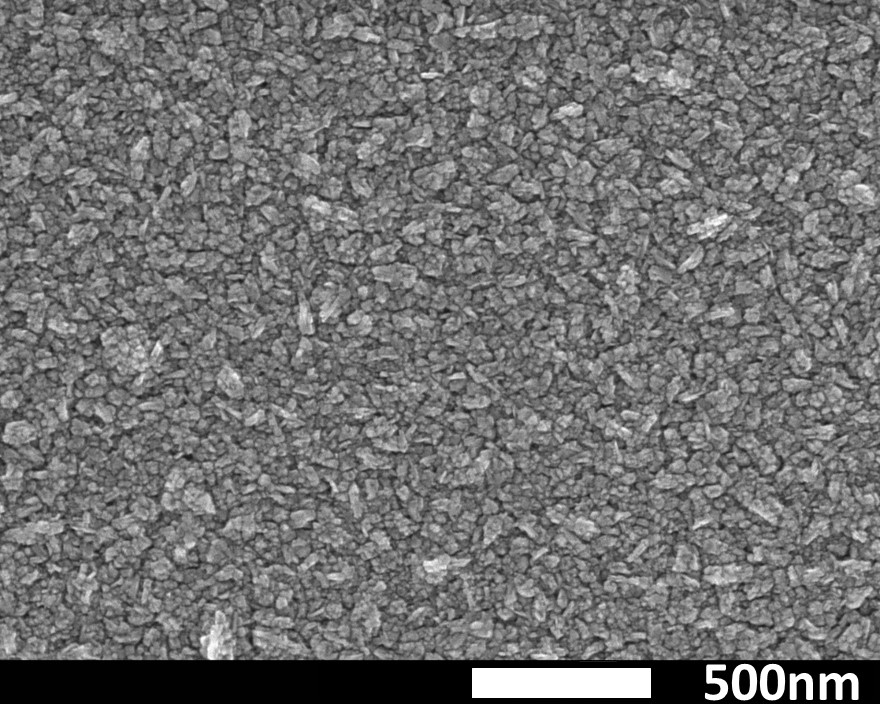} 
        \caption{Top view}
    \end{subfigure}
    \begin{subfigure}{0.3\textwidth}
        \includegraphics[width=\linewidth]{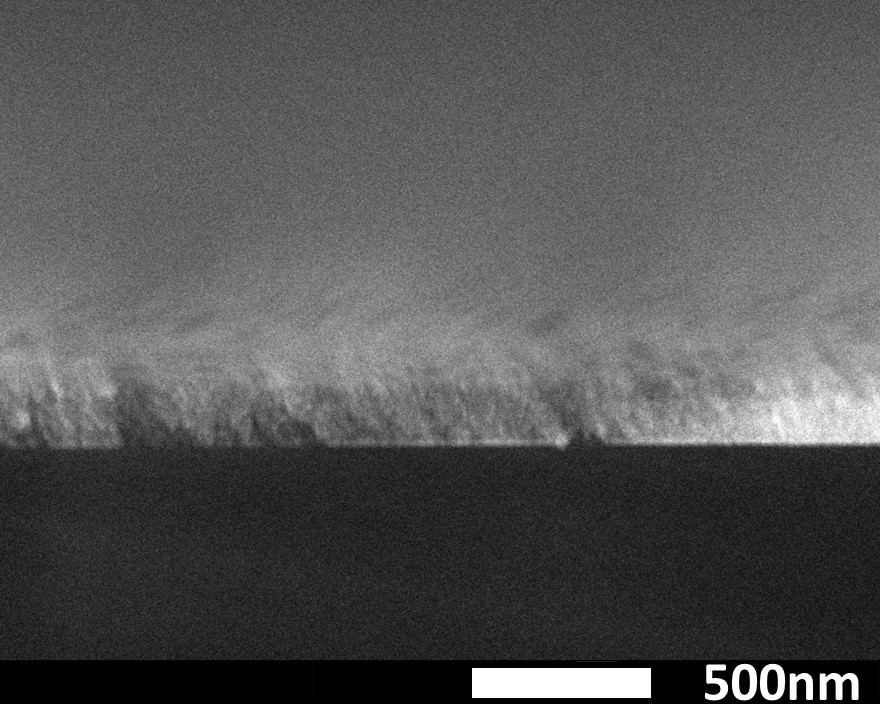}
        \caption{Cross-section}
    \end{subfigure}
    \begin{subfigure}{0.3\textwidth}
        \includegraphics[width=\linewidth]{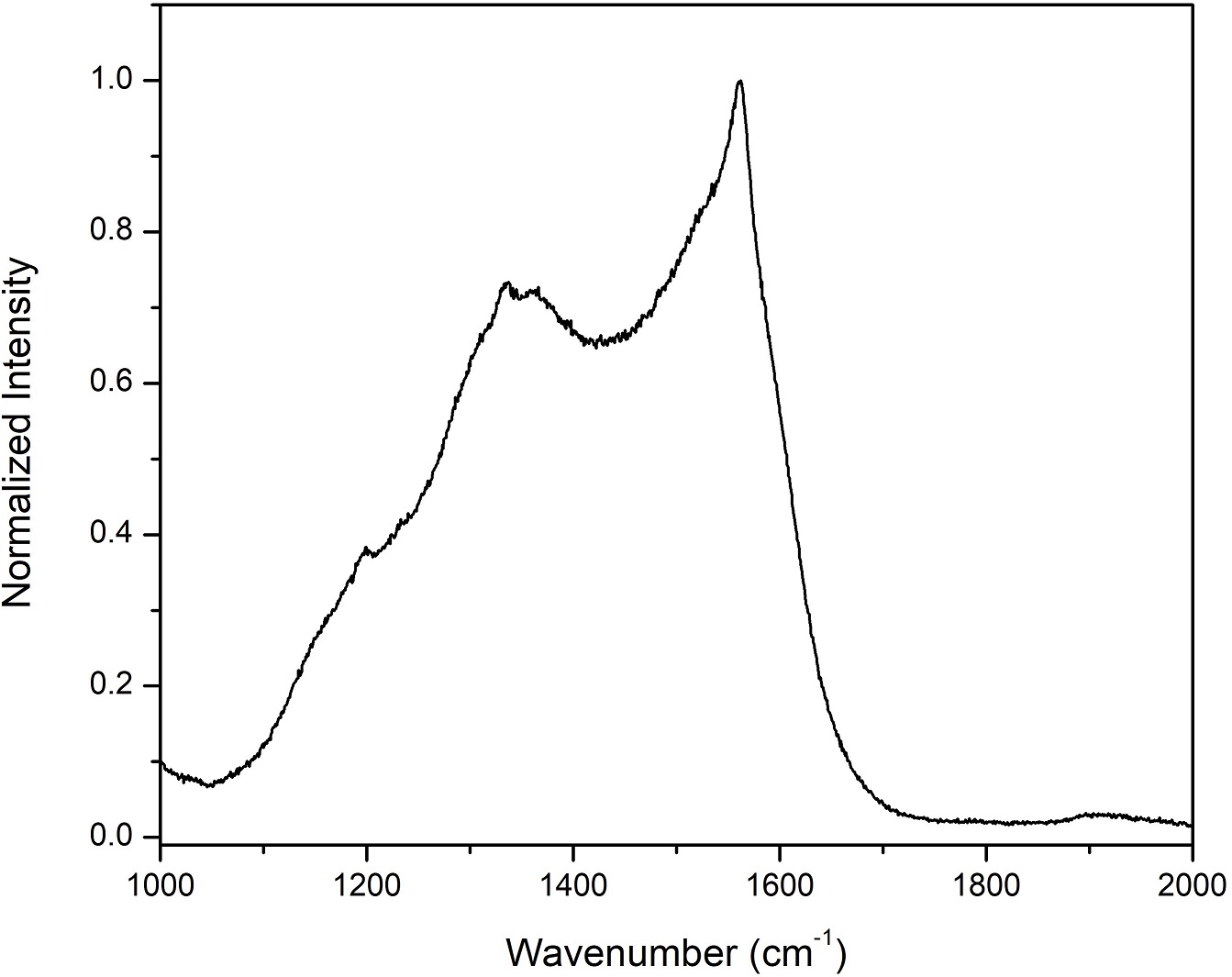}
        \caption{Raman spectrum}
    \end{subfigure}
    \vspace{-0pt}
\caption{SEM and Raman spectrum of sample:~SD1, grown with 20$\%$ N$_\text{2}$}
\label{fig:SD1}
\vspace{-0pt}

    \begin{subfigure}{0.3\textwidth}
        \includegraphics[width=\linewidth]{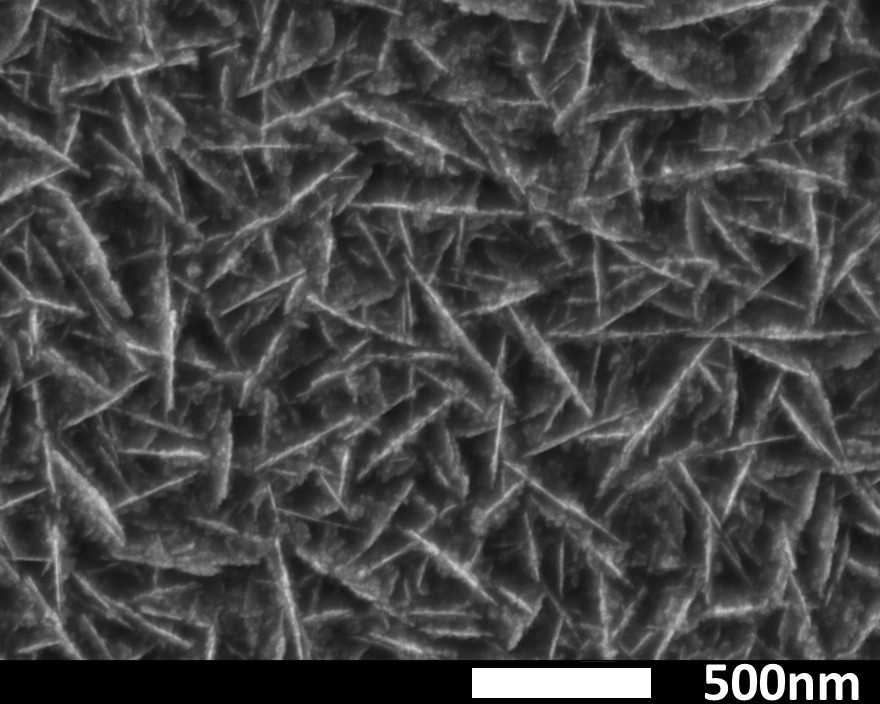} 
        \caption{Top view}
    \end{subfigure}
    \begin{subfigure}{0.3\textwidth}
        \includegraphics[width=\linewidth]{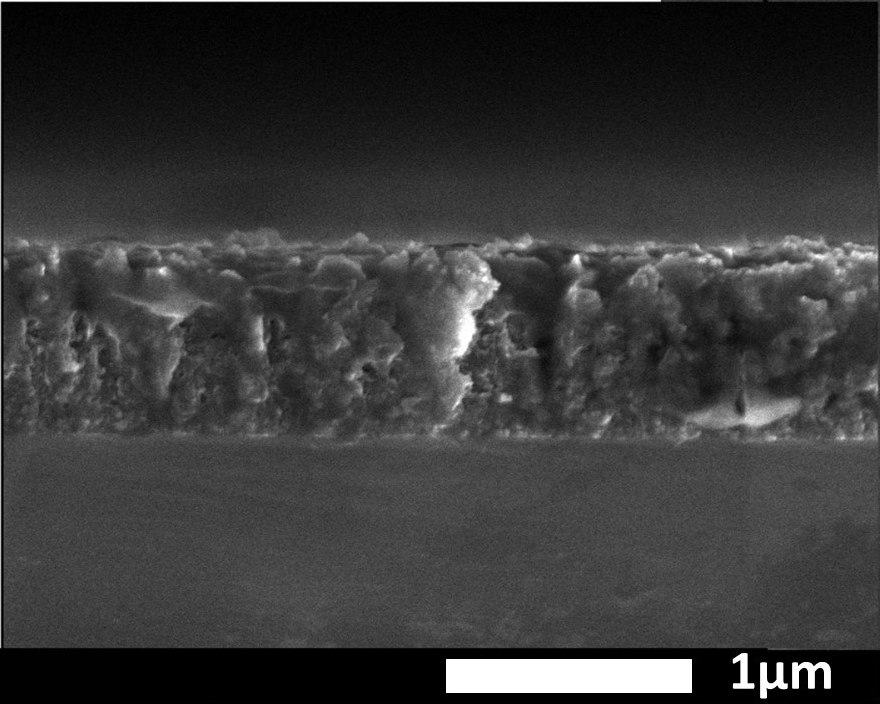}
        \caption{Cross-section}
    \end{subfigure}
    \begin{subfigure}{0.3\textwidth}
        \includegraphics[width=\linewidth]{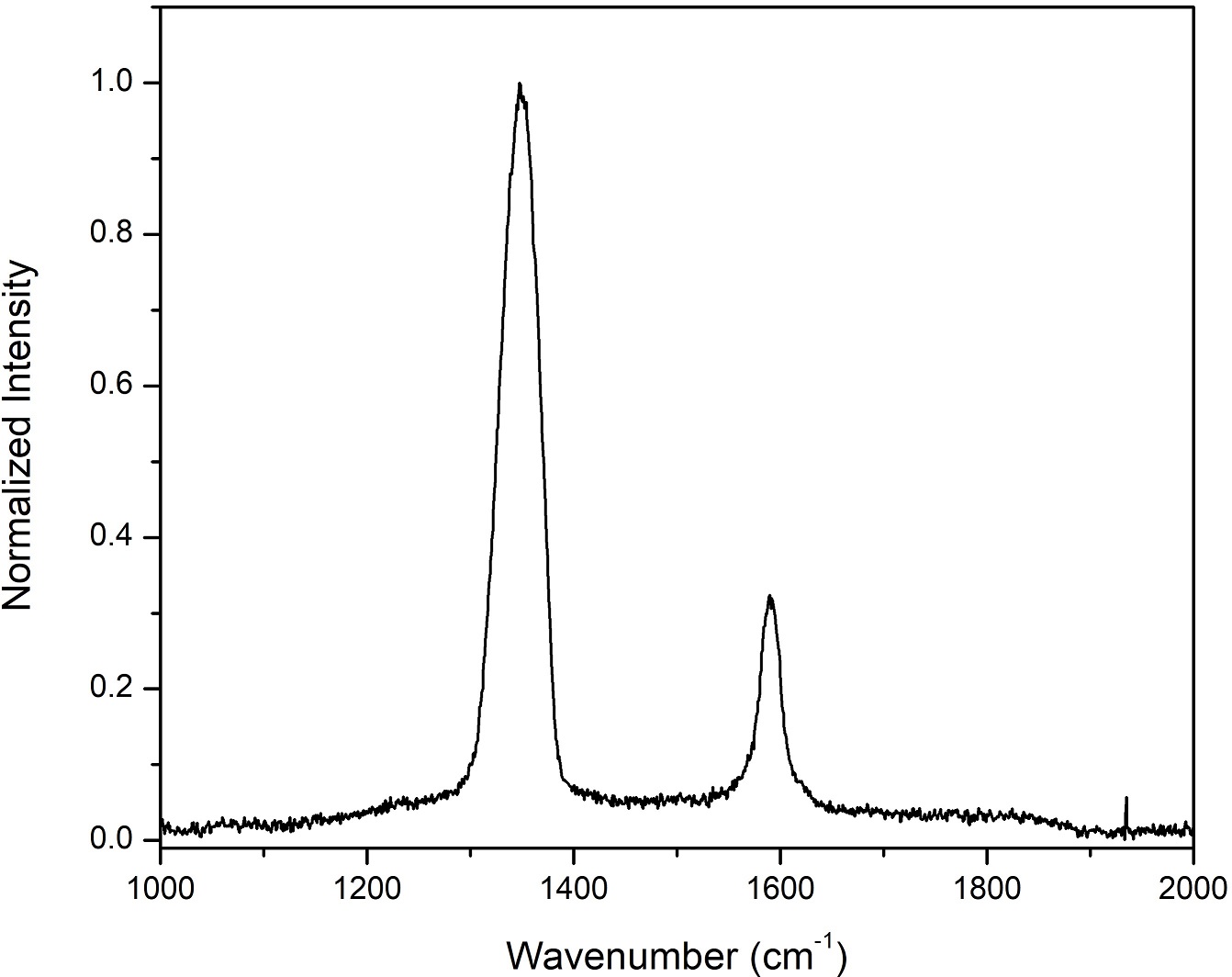}
        \caption{Raman Spectrum}
    \end{subfigure}
    \vspace{-0pt}
\caption{SEM and Raman spectrum of sample:~SB9, grown at high T$_\text{d}$}
\label{fig:SB9}
\vspace{-0pt}
\end{figure*} 

\subsubsection{Changes in surface morphology with varying T$_\text{d}$}
Comparison of the morphology of films grown at a particular nitrogen concentration shows a transition from UNCD to nanocrystalline graphite with increasing $T_d$. Films grown at lower $T_d$ as shown in Fig.~\ref{fig:SEM_N2_0}(a), \ref{fig:SEM_N2_5}(a), \ref{fig:SEM_N2_10}(a) and \ref{fig:SEM_N2_20}(a) exhibit densely packed nanometer sized diamond grains forming a continuous film and thus, confirming the growth of UNCD films. At higher deposition temperatures, it is seen that the amount of deposited graphite increases. Fig.~\ref{fig:SEM_N2_0}(b), \ref{fig:SEM_N2_5}(b), \ref{fig:SEM_N2_10}(b) and \ref{fig:SEM_N2_20}(b) show the appearance of small voids that grow as $T_d$ increases. These voids finally take form of ordered nano-platelets as seen in Fig.~\ref{fig:SEM_N2_0}(c), \ref{fig:SEM_N2_5}(c), \ref{fig:SEM_N2_10}(c) and \ref{fig:SEM_N2_20}(c). This is in accordance with the transition in morphology from densely packed nano-diamond film to platelet packed semi-void nano-graphite with increasing $T_d$ reported earlier \cite{Field_emission_properties_of_nanocrystalline_chemically_vapor_deposited_diamond_films}. Therefore, the surface morphology results are consistent with the Raman results.

\subsubsection{Changes in surface morphology with increasing N$_\text{2}$ concentration}
Films grown at lower $T_d$ reveal a sizable change in the grain size as nitrogen concentration is varied. For low N$_2$ concentration (Fig.~\ref{fig:SEM_N2_0}(a), \ref{fig:SEM_N2_5}(a) for 0\%, 5\% N$_2$ respectively), films are small grained. Fig.~\ref{fig:SEM_N2_10}(a) shows a slight increase in the grain size for 10\% N$_2$ film. However, grain size increases significantly and the (N)UNCD of SD series structure become coarse representing the strong influence of high nitrogen content (20\% here) on $sp^3$ phase growth, as seen from Fig.~\ref{fig:SEM_N2_20}(a). These results are in accordance with previously reported changes in grain size with varying nitrogen concentration \cite{Bhattacharyya2001,Birrell2002}. For the case of SD series, increase in $T_d$ additionally makes the grain size larger such that diamond lattice faceting becomes visible. This is consistent with the Raman results presented in Fig.~\ref{fig:Raman_N2_20} that illustrated the diamond D band (1333 cm$^{-1}$) resolved from the defect D band (1350 cm$^{-1}$).

\subsubsection{Cross section analysis}
To obtain more insight into the difference between closely grain packed and platelet packed films, cross-sectional analysis was employed, see Figs.~\ref{fig:SB4}, \ref{fig:SD1} and \ref{fig:SB9}. It is seen that the films grown at lower $T_d$ show no identifiable columnar structure for both low and high nitrogen concentration samples -- this is a canonical UNCD pattern due to high re-nucleation rate \cite{Gruen_1999}. More specifically, the cross-section of 5\% N$_2$ film in Fig.~\ref{fig:SB4}(b) shows smooth fracture cross-section of a highly dense continuous film with no columnar growth. The cross-section of 20$\%$ N$_\text{2}$ film in Fig.~\ref{fig:SD1}(b) also shows a highly dense film resulting from high renuclation rate along with discernable grains due to larger crystallite size as seen in Fig.~\ref{fig:SD1}(a).

Contrastingly, Fig.~\ref{fig:SB9}(b) shows the cross-sectional view of highly graphitized film (as interpreted from Raman) grown at a high $T_d$ of 1229 K: the platelet observed in the top-view imaging thread through the entire film, from the substrate to the surface.

\subsection{Graphitization}
\begin{figure}
\centering
\includegraphics[width =0.48\textwidth]{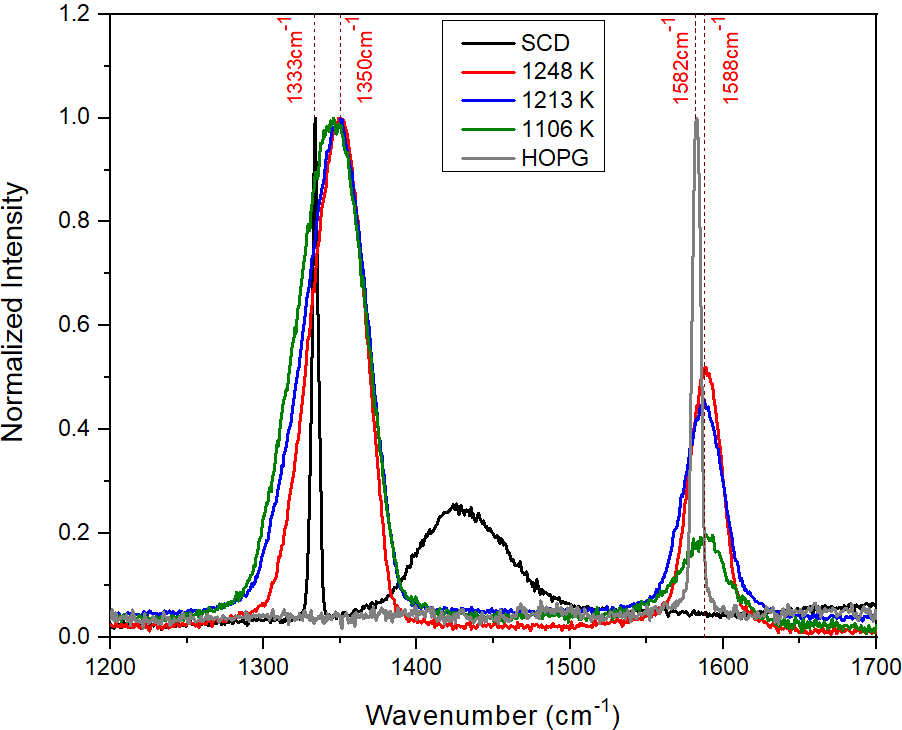}
\caption{Comparison of Raman spectra of highly graphitized films with single crytsal diamond and highly oriented pyrolytic graphite} 
\label{fig:Graphitization}
\end{figure}

The significant change in the peak positions and width of the D and G bands has been noted for all the four series of films. It was interpreted to be due to material transition from nano-sized diamond grains surrounded by amorphous $sp^2$ carbon liner to nanocrystalline graphite. This interpretation was further supported by top-view and cross-sectional SEM imaging. To even further corroborate this argument and confirm the process of graphitization in the films, a comparison with single crystal diamond (SCD), a pure $sp^3$ case, and highly oriented pyrolytic graphite (HOPG), a pure $sp^2$ case, was drawn. Raman spectra (Fig.~\ref{fig:Graphitization}) of SCD with peak at 1333 cm$^{-1}$ and HOPG with peak at 1582 cm$^{-1}$ are compared to a 0 \% N$_2$ UNCD grown at three highest temperatures, 1156, 1213 and 1248 K. As was evident from Fig.~\ref{fig:Raman_N2_0}, the intensity of G band peak at 1550 cm$^\text{-1}$ of the 0$\%$ N$_\text{2}$ films keeps decreasing owing to the decrease in the amorphous content in the film along with the shift of the peak position from 1550 cm$^\text{-1}$ to 1588 cm$^\text{-1}$. When the peak has completely shifted to 1588 cm$^\text{-1}$, it does not change its position any further with an increase in $T_d$ beyond 1156 K and starts re-intensifying again as shown now in Fig.~\ref{fig:Graphitization}. The increase in the 1588 cm$^\text{-1}$ peak shows an increase in the amount of crystalline $sp^2$ phase. The absence of the diamond peak feature within the D band and the presence of the G band in the spectra of 0\% N$_2$ films (as opposed to single G band spectrum of HOPG) confirms that now the material grows completely as nanocrystalline graphite. The increase in the crystallinity of $sp^2$ phase is also evident from Fig.~\ref{fig:Raman_N2_5} for the 5\% N$_2$ films where G band peak completely shifts to 1590 cm$^{-1}$ at 1163 K and starts intensifying for higher $T_d$.

A similar material transition was also observed in our study of glow discharge plasma during field emission from UNCD cathodes \cite{Our_Paper} where data from SEM and Raman showed the conversion of nano-diamond to nano-graphite caused by the electron emission self-induced heating of the emitter to 2000--4000 K. 

\begin{figure}[H]
\centering
\includegraphics[width =0.48\textwidth]{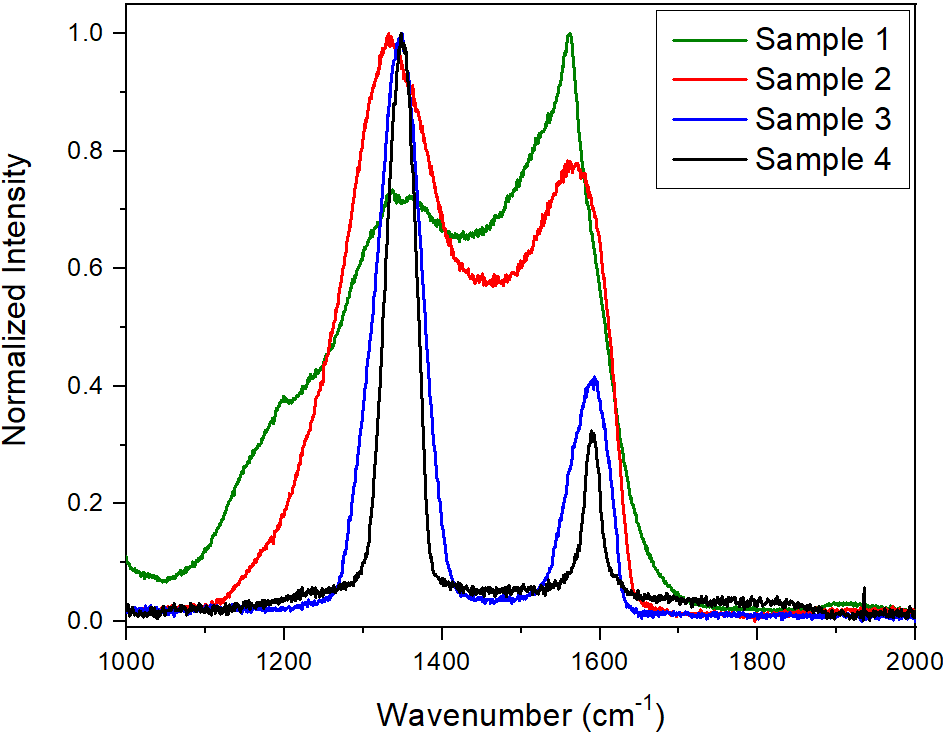}
\caption{Corresponding Raman spectra of the samples used for thickness comparison in Table V} 
\label{fig:thickness_raman}
\end{figure}

The most direct confirmation of graphitization comes from mass density evaluation. We compare the actual thickness measurement from the cross-section SEM imaging ($t_{SEM}$) with indirect thickness estimation from the weight difference before and after growth ($t_{weight}$) using the density of pure diamond ($\rho_{dia}$=3.51 g/cm$^3$) and graphite ($\rho_{gra}$=2.27 g/cm$^3$). As seen from Table.~\ref{table:thickness}, $t_{SEM}$ values are in close match with $t_{weight}$ calculated using $\rho_{dia}$ for samples grown at low $T_d$ (Samples 1 and 2), thus confirming they largely retain the structural property of diamond (Raman spectra of these samples are presented in Fig.~\ref{fig:thickness_raman}). For high $T_d$ though, $t_{weight}$ only matches $t_{SEM}$ if $\rho_{gra}$ is used for calculation. Thus, Samples 3 and 4 (Raman spectra are in Fig.~\ref{fig:thickness_raman}) largely retain the structural property of graphite.

\begin{table}[H]
\centering
\resizebox{0.4\textwidth}{!}{%
\begin{tabular}{|c|c|c|c|}
\hline
\textbf{Sample} & \textbf{\begin{tabular}[c]{@{}l@{}}$t_{SEM}$ (nm)\end{tabular}} & \textbf{\begin{tabular}[c]{@{}l@{}} $t_{weight}$ (nm)\\ using $\rho_{dia}$ \end{tabular}} & \textbf{\begin{tabular}[c]{@{}l@{}}t$_\text{weight}$ (nm) \\ using $\rho_{gra}$ \end{tabular}}\\ \hline
Sample 1 & 156 & 194 & 300 \\ \hline
Sample 2 & 764 & 636 & 983 \\ \hline
Sample 3 & 742 & 517 & 799 \\ \hline
Sample 4 & 830 & 532 & 822 \\ \hline
\end{tabular}%
}
\caption{Comparison of thickness values obtained from SEM and weight density measurements}\label{table:thickness}
\end{table}

\begin{figure}[H]
\centering
\includegraphics[width =0.48\textwidth]{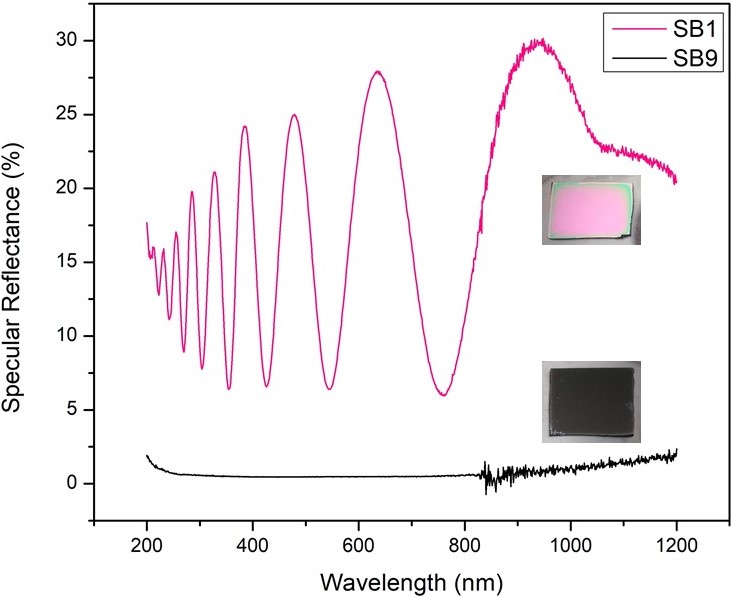}
\caption{Comparison of reflectance spectra of UNCD and highly graphitized film} 
\label{fig:UV_vis}
\end{figure}

A  comparison of the reflectance spectrum obtained from UV-vis spectroscopy is shown in Fig.~\ref{fig:UV_vis}. High reflectance and strong interference pattern was exhibited by the bright pink sample SB1 deposited at 1043 K while low reflectance and no interference was observed for the highly graphitized sample SB9 grown at 1229 K. Both, mass density and optical analysis provide unambiguous proof of the fundamental material transition from UNCD to nanocrystalline graphite.

\subsection{Revisiting resistivity effects and attempting to redefine UNCD}

\begin{figure}
\centering
\includegraphics[width =0.48\textwidth]{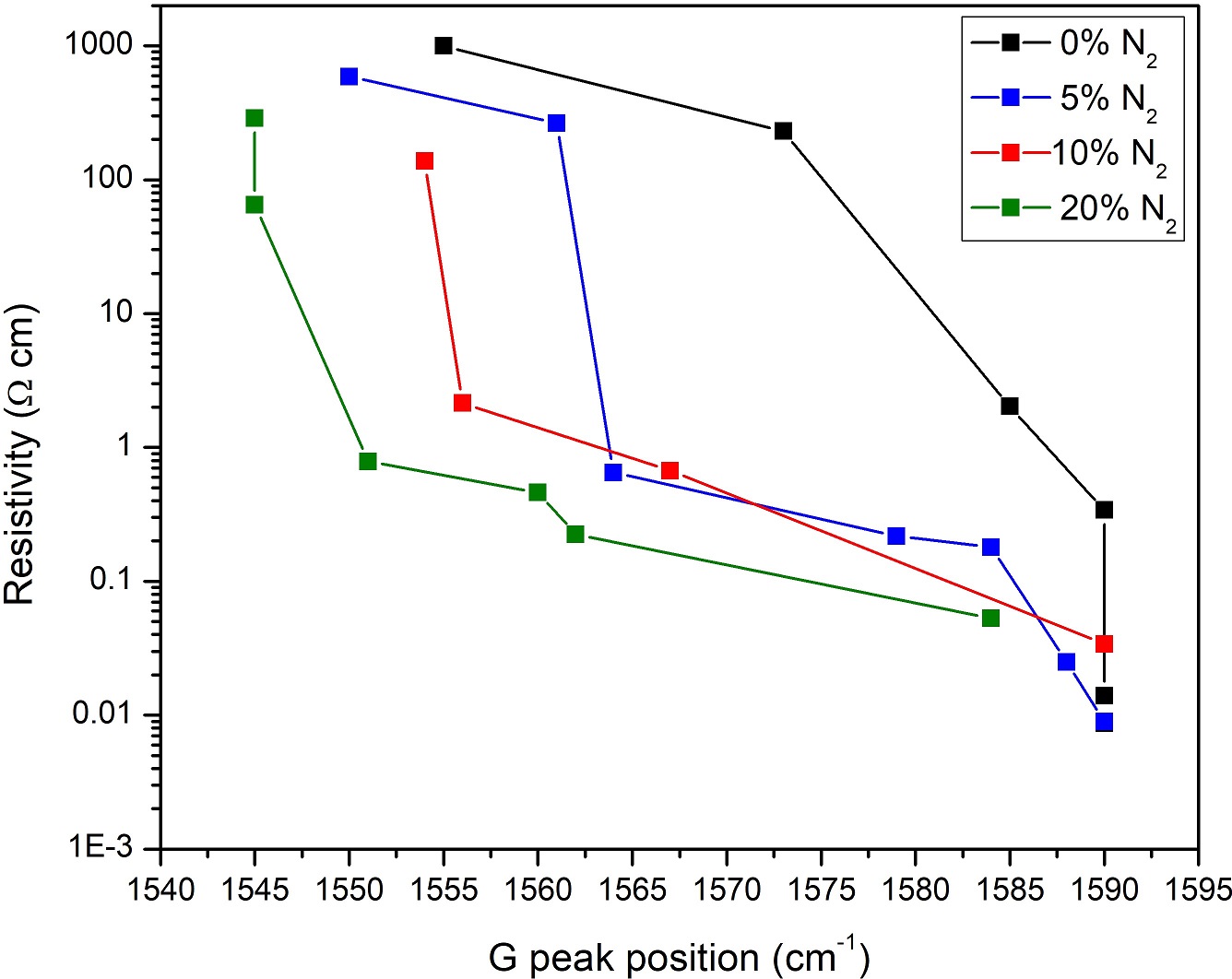}
\caption{Resistivity as a function of G peak position} 
\label{fig:R_vs_Gpeak}
\end{figure} 

Having confirmed the UNCD--nanographite transformation, a more universal chart behind resistivity effect, that depends not only on N$_2$ content but also on $T_d$, can be drawn. It is proposed to plot resistivity as a function of G band peak position, as examples in Fig.~\ref{fig:R_vs_Gpeak}, with G peak position bound between 1550 cm$^{-1}$ to 1590 cm$^{-1}$. Four samples series are plotted. A few main features can be highlighted as follows:

\noindent 1) The resistivity of series SD (20\% N$_2$) has immediate and strong response to the G peak position: slight improvement in GB crystallinity (G band shifts by +5 cm$^{-1}$) drops resistivity by over 2 orders of magnitude;

\noindent 2) Unlike series SD, series SA (0\% N$_2$) is largely insensitive to the G band position (GB crystallinity) until after 1575 cm$^{-1}$. 

\noindent 3) As $T_d$ reaches the highest values when all the films (regardless of N$_2$ content) turn into crystalline nanographite (G band is centered at 1590 cm$^{-1}$), it is not surprising to see that all the four film series converge to the same resistivity of $\sim$$10^{-2}$~$\Omega\cdot$cm at the same G peak position centered at 1590 cm$^{-1}$.

As was demonstrated in Figs.~\ref{fig:1098K} and \ref{fig:1126K}, the increase of plasma nitrogen content simultaneously increases grain size and GB size. This is consistent with previous transmission electron microscopy results \cite{Birrell2002, Bhattacharyya2001} that found changing N$_2$ from 0\% to 20\% inflated the GB width from 0.5 to 2.2 nm and diamond grain size from 4 to 16 nm. If so, the following physical mechanism is proposed to interpret the result of Fig.~\ref{fig:R_vs_Gpeak}. Higher N$_2$ content creates larger grain boundaries: it means larger density of state (DOS) in the $\pi$ and $\pi^*$ bands that reside inside the fundamental band gap of diamond \cite{Nesladek1996, Zapol2001, Oksana2019}, plus it increases the physical connectivity between touching GBs (contact area). Thus, at high N$_2$ concentration the large amount of charge is readily available and the cross section area of wired/interconnected GBs is large. Even so, at the lowest $T_d$, attempted in this work, the resistivity remains very high -- this is because the charge carrier mean free path is small and so is the mobility due to amorphous structure in GBs. The role of increasing $T_d$ is two-fold: (1) it keeps increasing the number of states in the $\pi$ and $\pi^*$ bands and (2) improves GB crystallinity yielding larger mean free path i.e. higher mobility.
Contrastingly, 0\% films have low connectivity and low DOS in the the $\pi$ and $\pi^*$ bands at low $T_d$, and can only develop those metrics (plus improved mobility owing to improved crystallinity) if the synthetis temperature is increased. This is why the SD series is more prone to the G band peak position, having immediate response to the quality of $sp^2$ GBs, than the other series.

Tracing now the D band position in our Raman spectra data sets, one more universal conclusion could be made. The film could be referred to as UNCD as long as it has some diamond grain material in it. As was illustrated, presence of diamond material is signified by the presence of the D band centered at 1333 cm$^{-1}$ peak, or at least an asymmetric left-hand shoulder of the D band extending toward 1333 cm$^{-1}$. The material fully transforms to nanocrystalline graphite when the D band completely shifts to nearby wavenumber 1350 cm$^{-1}$ (at this point the D band is symmetric) and therefore no longer probes/detects diamond $sp^3$ content in the film but rather represents nanographite defect condition.

\section{Conclusion}
Effects of nitrogen incorporation (via changing the N$_2$ content in the synthetic plasma from 0 to 20 \%) and the deposition temperature $T_d$ (via changing the total plasma pressure and input rf power from 1000 to 1300 K) on (N)UNCD resistivity were systematically investigated using a set of 27 samples.

Based on Raman spectroscopy (additionally supported by physical structure studies by SEM, spectrophotometry and weighing), charge transport can be explained based on two competitive processes. Addition of N$_2$ increases the $sp^2$-bonded carbon comprising GBs, which yields enhanced spatial connectivity of grain boundaries and enhanced overlap between $\pi$ and $\pi^*$ states in the energy space. Together, these could make charge transport more facile causing resistivity to drop. In reality, resistivity does not drop until an elevated $T_d$ improves crystallinity (enhancing mobility) of GBs because incorporated nitrogen amorphizes them. 
Larger the N$_2$ content, larger the $T_d$ required to recover to the same resistivity. Therefore, there is a thermophysicochemical process that governs the structure and kinetics of the GB formation that is dependent on flow rates, total pressure and input rf power -- this process is a function of the reactor design with $T_d$ being a byproduct of the reactor settings/design. It is proposed to represent resistivity as function of the Raman G band peak position, one of the two main characteristic bands in UNCD (as well as other polycrystalline diamond and graphite materials) storing information about $sp^2$ grain boundary structure. This representation is not contingent on the reactor design and allows for relating the resistivity and structural quality of the final material product.

Practically speaking, this approach outlines a path of engineering resistivity of UNCD between $10^3-10^{-2}$ $\Omega$ cm even when containing no nitrogen. The downside of synthesizing UNCD at 0 \% $N_2$ is that achieving resistivity of $10^{-1}-10^{-2}$ $\Omega$ cm appealing for many applications requires $T_d$ in excess of 1150 K (in our reactor design): at this point the material retains little or no signature of UNCD and is being transformed into nanographite. In contrast, enhanced connectivity between GBs in N$_2$ rich samples allows for achieving $10^{-1}-10^{-2}$ $\Omega$ cm while maintaining high diamond fraction in UNCD, thanks to the slowed kinetics of the nanodiamond-to-nanographite phase transition taking place at $T_d$ over 1250 K (in our reactor design).

\begin{acknowledgments}
The authors would like to thank James E. Butler, Tim Hogan and Tim Grotjohn for useful discussions. TN and SVB were funded by the College of
Engineering, Michigan State University, under the Global
Impact Initiative.
\end{acknowledgments}

\bibliography{Manuscript}

\begin{thebibliography}{30}%
\makeatletter
\providecommand \@ifxundefined [1]{%
 \@ifx{#1\undefined}
}%
\providecommand \@ifnum [1]{%
 \ifnum #1\expandafter \@firstoftwo
 \else \expandafter \@secondoftwo
 \fi
}%
\providecommand \@ifx [1]{%
 \ifx #1\expandafter \@firstoftwo
 \else \expandafter \@secondoftwo
 \fi
}%
\providecommand \natexlab [1]{#1}%
\providecommand \enquote  [1]{``#1''}%
\providecommand \bibnamefont  [1]{#1}%
\providecommand \bibfnamefont [1]{#1}%
\providecommand \citenamefont [1]{#1}%
\providecommand \href@noop [0]{\@secondoftwo}%
\providecommand \href [0]{\begingroup \@sanitize@url \@href}%
\providecommand \@href[1]{\@@startlink{#1}\@@href}%
\providecommand \@@href[1]{\endgroup#1\@@endlink}%
\providecommand \@sanitize@url [0]{\catcode `\\12\catcode `\$12\catcode
  `\&12\catcode `\#12\catcode `\^12\catcode `\_12\catcode `\%12\relax}%
\providecommand \@@startlink[1]{}%
\providecommand \@@endlink[0]{}%
\providecommand \url  [0]{\begingroup\@sanitize@url \@url }%
\providecommand \@url [1]{\endgroup\@href {#1}{\urlprefix }}%
\providecommand \urlprefix  [0]{URL }%
\providecommand \Eprint [0]{\href }%
\providecommand \doibase [0]{https://doi.org/}%
\providecommand \selectlanguage [0]{\@gobble}%
\providecommand \bibinfo  [0]{\@secondoftwo}%
\providecommand \bibfield  [0]{\@secondoftwo}%
\providecommand \translation [1]{[#1]}%
\providecommand \BibitemOpen [0]{}%
\providecommand \bibitemStop [0]{}%
\providecommand \bibitemNoStop [0]{.\EOS\space}%
\providecommand \EOS [0]{\spacefactor3000\relax}%
\providecommand \BibitemShut  [1]{\csname bibitem#1\endcsname}%
\let\auto@bib@innerbib\@empty
\bibitem [{\citenamefont {Landstrass}\ and\ \citenamefont
  {Ravi}(1989)}]{Landstrass1989}%
  \BibitemOpen
  \bibfield  {author} {\bibinfo {author} {\bibfnamefont {M.~I.}\ \bibnamefont
  {Landstrass}}\ and\ \bibinfo {author} {\bibfnamefont {K.~V.}\ \bibnamefont
  {Ravi}},\ }\bibfield  {title} {\bibinfo {title} {Resistivity of chemical
  vapor deposited diamond films},\ }\href@noop {} {\bibfield  {journal}
  {\bibinfo  {journal} {Applied Physics Letters}\ }\textbf {\bibinfo {volume}
  {55}},\ \bibinfo {pages} {975} (\bibinfo {year} {1989})}\BibitemShut
  {NoStop}%
\bibitem [{\citenamefont {Prins}(2000)}]{Prins2000}%
  \BibitemOpen
  \bibfield  {author} {\bibinfo {author} {\bibfnamefont {J.~F.}\ \bibnamefont
  {Prins}},\ }\bibfield  {title} {\bibinfo {title} {n-type semiconducting
  diamond by means of oxygen-ion implantation},\ }\href@noop {} {\bibfield
  {journal} {\bibinfo  {journal} {Phys. Rev. B}\ }\textbf {\bibinfo {volume}
  {61}},\ \bibinfo {pages} {7191} (\bibinfo {year} {2000})}\BibitemShut
  {NoStop}%
\bibitem [{\citenamefont {Sakaguchi}\ \emph {et~al.}(1999)\citenamefont
  {Sakaguchi}, \citenamefont {N.-Gamo}, \citenamefont {Kikuchi}, \citenamefont
  {Yasu}, \citenamefont {Haneda}, \citenamefont {Suzuki},\ and\ \citenamefont
  {Ando}}]{Sakaguchi1999}%
  \BibitemOpen
  \bibfield  {author} {\bibinfo {author} {\bibfnamefont {I.}~\bibnamefont
  {Sakaguchi}}, \bibinfo {author} {\bibfnamefont {M.}~\bibnamefont {N.-Gamo}},
  \bibinfo {author} {\bibfnamefont {Y.}~\bibnamefont {Kikuchi}}, \bibinfo
  {author} {\bibfnamefont {E.}~\bibnamefont {Yasu}}, \bibinfo {author}
  {\bibfnamefont {H.}~\bibnamefont {Haneda}}, \bibinfo {author} {\bibfnamefont
  {T.}~\bibnamefont {Suzuki}},\ and\ \bibinfo {author} {\bibfnamefont
  {T.}~\bibnamefont {Ando}},\ }\bibfield  {title} {\bibinfo {title} {Sulfur: A
  donor dopant for n-type diamond semiconductors},\ }\href@noop {} {\bibfield
  {journal} {\bibinfo  {journal} {Phys. Rev. B}\ }\textbf {\bibinfo {volume}
  {60}},\ \bibinfo {pages} {R2139} (\bibinfo {year} {1999})}\BibitemShut
  {NoStop}%
\bibitem [{\citenamefont {Grot}\ \emph {et~al.}(1991)\citenamefont {Grot},
  \citenamefont {Hatfield}, \citenamefont {Gildenblat}, \citenamefont
  {Badzian},\ and\ \citenamefont {Badzian}}]{Grot1991}%
  \BibitemOpen
  \bibfield  {author} {\bibinfo {author} {\bibfnamefont {S.~A.}\ \bibnamefont
  {Grot}}, \bibinfo {author} {\bibfnamefont {C.~W.}\ \bibnamefont {Hatfield}},
  \bibinfo {author} {\bibfnamefont {G.~S.}\ \bibnamefont {Gildenblat}},
  \bibinfo {author} {\bibfnamefont {A.~R.}\ \bibnamefont {Badzian}},\ and\
  \bibinfo {author} {\bibfnamefont {T.}~\bibnamefont {Badzian}},\ }\bibfield
  {title} {\bibinfo {title} {Electrical properties of selectively grown
  homoepitaxial diamond films},\ }\href@noop {} {\bibfield  {journal} {\bibinfo
   {journal} {Applied Physics Letters}\ }\textbf {\bibinfo {volume} {58}},\
  \bibinfo {pages} {1542} (\bibinfo {year} {1991})}\BibitemShut {NoStop}%
\bibitem [{\citenamefont {Williams}\ \emph {et~al.}(2001)\citenamefont
  {Williams}, \citenamefont {Whitfield}, \citenamefont {Jackman}, \citenamefont
  {Foord}, \citenamefont {Butler},\ and\ \citenamefont {Nebel}}]{Williams2001}%
  \BibitemOpen
  \bibfield  {author} {\bibinfo {author} {\bibfnamefont {O.~A.}\ \bibnamefont
  {Williams}}, \bibinfo {author} {\bibfnamefont {M.~D.}\ \bibnamefont
  {Whitfield}}, \bibinfo {author} {\bibfnamefont {R.~B.}\ \bibnamefont
  {Jackman}}, \bibinfo {author} {\bibfnamefont {J.~S.}\ \bibnamefont {Foord}},
  \bibinfo {author} {\bibfnamefont {J.~E.}\ \bibnamefont {Butler}},\ and\
  \bibinfo {author} {\bibfnamefont {C.~E.}\ \bibnamefont {Nebel}},\ }\bibfield
  {title} {\bibinfo {title} {Formation of shallow acceptor states in the
  surface region of thin film diamond},\ }\href@noop {} {\bibfield  {journal}
  {\bibinfo  {journal} {Applied Physics Letters}\ }\textbf {\bibinfo {volume}
  {78}},\ \bibinfo {pages} {3460} (\bibinfo {year} {2001})}\BibitemShut
  {NoStop}%
\bibitem [{\citenamefont {Koizumi}\ \emph {et~al.}(2000)\citenamefont
  {Koizumi}, \citenamefont {Teraji},\ and\ \citenamefont
  {Kanda}}]{KOIZUMI20009}%
  \BibitemOpen
  \bibfield  {author} {\bibinfo {author} {\bibfnamefont {S.}~\bibnamefont
  {Koizumi}}, \bibinfo {author} {\bibfnamefont {T.}~\bibnamefont {Teraji}},\
  and\ \bibinfo {author} {\bibfnamefont {H.}~\bibnamefont {Kanda}},\ }\bibfield
   {title} {\bibinfo {title} {Phosphorus-doped chemical vapor deposition of
  diamond},\ }\href@noop {} {\bibfield  {journal} {\bibinfo  {journal} {Diamond
  and Related Materials}\ }\textbf {\bibinfo {volume} {9}},\ \bibinfo {pages}
  {935 } (\bibinfo {year} {2000})}\BibitemShut {NoStop}%
\bibitem [{\citenamefont {Kato}\ \emph {et~al.}(2005)\citenamefont {Kato},
  \citenamefont {Yamasaki},\ and\ \citenamefont {Okushi}}]{Kato2005}%
  \BibitemOpen
  \bibfield  {author} {\bibinfo {author} {\bibfnamefont {H.}~\bibnamefont
  {Kato}}, \bibinfo {author} {\bibfnamefont {S.}~\bibnamefont {Yamasaki}},\
  and\ \bibinfo {author} {\bibfnamefont {H.}~\bibnamefont {Okushi}},\
  }\bibfield  {title} {\bibinfo {title} {n-type doping of (001)-oriented
  single-crystalline diamond by phosphorus},\ }\href@noop {} {\bibfield
  {journal} {\bibinfo  {journal} {Applied Physics Letters}\ }\textbf {\bibinfo
  {volume} {86}},\ \bibinfo {pages} {222111} (\bibinfo {year}
  {2005})}\BibitemShut {NoStop}%
\bibitem [{\citenamefont {Auciello}\ \emph {et~al.}(2014)\citenamefont
  {Auciello}, \citenamefont {Gurman}, \citenamefont {Guglielmotti},
  \citenamefont {Olmedo}, \citenamefont {Berra},\ and\ \citenamefont
  {Saravia}}]{auciello_gurman_guglielmotti_olmedo_berra_saravia_2014}%
  \BibitemOpen
  \bibfield  {author} {\bibinfo {author} {\bibfnamefont {O.}~\bibnamefont
  {Auciello}}, \bibinfo {author} {\bibfnamefont {P.}~\bibnamefont {Gurman}},
  \bibinfo {author} {\bibfnamefont {M.~B.}\ \bibnamefont {Guglielmotti}},
  \bibinfo {author} {\bibfnamefont {D.~G.}\ \bibnamefont {Olmedo}}, \bibinfo
  {author} {\bibfnamefont {A.}~\bibnamefont {Berra}},\ and\ \bibinfo {author}
  {\bibfnamefont {M.~J.}\ \bibnamefont {Saravia}},\ }\bibfield  {title}
  {\bibinfo {title} {Biocompatible ultrananocrystalline diamond coatings for
  implantable medical devices},\ }\href {https://doi.org/10.1557/mrs.2014.134}
  {\bibfield  {journal} {\bibinfo  {journal} {MRS Bulletin}\ }\textbf {\bibinfo
  {volume} {39}},\ \bibinfo {pages} {621–629} (\bibinfo {year}
  {2014})}\BibitemShut {NoStop}%
\bibitem [{\citenamefont {Auciello}\ and\ \citenamefont
  {Sumant}(2010)}]{article}%
  \BibitemOpen
  \bibfield  {author} {\bibinfo {author} {\bibfnamefont {O.}~\bibnamefont
  {Auciello}}\ and\ \bibinfo {author} {\bibfnamefont {A.}~\bibnamefont
  {Sumant}},\ }\bibfield  {title} {\bibinfo {title} {Status review of the
  science and technology of ultrananocrystalline diamond (uncd (tm)) films and
  application to multifunctional devices},\ }\href
  {https://doi.org/10.1016/j.diamond.2010.03.015} {\bibfield  {journal}
  {\bibinfo  {journal} {Diamond and Related Materials}\ }\textbf {\bibinfo
  {volume} {19}},\ \bibinfo {pages} {699} (\bibinfo {year} {2010})}\BibitemShut
  {NoStop}%
\bibitem [{\citenamefont {Bajaj}\ \emph {et~al.}(2007)\citenamefont {Bajaj},
  \citenamefont {Akin}, \citenamefont {Gupta}, \citenamefont {Sherman},
  \citenamefont {Shi}, \citenamefont {Auciello},\ and\ \citenamefont
  {Bashir}}]{Bajaj2007}%
  \BibitemOpen
  \bibfield  {author} {\bibinfo {author} {\bibfnamefont {P.}~\bibnamefont
  {Bajaj}}, \bibinfo {author} {\bibfnamefont {D.}~\bibnamefont {Akin}},
  \bibinfo {author} {\bibfnamefont {A.}~\bibnamefont {Gupta}}, \bibinfo
  {author} {\bibfnamefont {D.}~\bibnamefont {Sherman}}, \bibinfo {author}
  {\bibfnamefont {B.}~\bibnamefont {Shi}}, \bibinfo {author} {\bibfnamefont
  {O.}~\bibnamefont {Auciello}},\ and\ \bibinfo {author} {\bibfnamefont
  {R.}~\bibnamefont {Bashir}},\ }\bibfield  {title} {\bibinfo {title}
  {Ultrananocrystalline diamond film as an optimal cell interface for
  biomedical applications},\ }\href {https://doi.org/10.1007/s10544-007-9090-2}
  {\bibfield  {journal} {\bibinfo  {journal} {Biomedical Microdevices}\
  }\textbf {\bibinfo {volume} {9}},\ \bibinfo {pages} {787} (\bibinfo {year}
  {2007})}\BibitemShut {NoStop}%
\bibitem [{\citenamefont {Xiao}\ \emph {et~al.}(2006)\citenamefont {Xiao},
  \citenamefont {Wang}, \citenamefont {Liu}, \citenamefont {Carlisle},
  \citenamefont {Mech}, \citenamefont {Greenberg}, \citenamefont {Guven},
  \citenamefont {Freda}, \citenamefont {Humayun}, \citenamefont {Weiland},\
  and\ \citenamefont {Auciello}}]{Xiao2006InVA}%
  \BibitemOpen
  \bibfield  {author} {\bibinfo {author} {\bibfnamefont {X.~M.}\ \bibnamefont
  {Xiao}}, \bibinfo {author} {\bibfnamefont {J.}~\bibnamefont {Wang}}, \bibinfo
  {author} {\bibfnamefont {C.}~\bibnamefont {Liu}}, \bibinfo {author}
  {\bibfnamefont {J.~A.}\ \bibnamefont {Carlisle}}, \bibinfo {author}
  {\bibfnamefont {B.~V.}\ \bibnamefont {Mech}}, \bibinfo {author}
  {\bibfnamefont {R.~G.}\ \bibnamefont {Greenberg}}, \bibinfo {author}
  {\bibfnamefont {D.~S.}\ \bibnamefont {Guven}}, \bibinfo {author}
  {\bibfnamefont {R.}~\bibnamefont {Freda}}, \bibinfo {author} {\bibfnamefont
  {M.~S.}\ \bibnamefont {Humayun}}, \bibinfo {author} {\bibfnamefont {J.~D.}\
  \bibnamefont {Weiland}},\ and\ \bibinfo {author} {\bibfnamefont {O.~H.}\
  \bibnamefont {Auciello}},\ }\bibfield  {title} {\bibinfo {title} {In vitro
  and in vivo evaluation of ultrananocrystalline diamond for coating of
  implantable retinal microchips.},\ }\href@noop {} {\bibfield  {journal}
  {\bibinfo  {journal} {Journal of biomedical materials research B}\ }\textbf
  {\bibinfo {volume} {77 2}},\ \bibinfo {pages} {273} (\bibinfo {year}
  {2006})}\BibitemShut {NoStop}%
\bibitem [{\citenamefont {Birrell}\ \emph {et~al.}(2002)\citenamefont
  {Birrell}, \citenamefont {Carlisle}, \citenamefont {Auciello}, \citenamefont
  {Gruen},\ and\ \citenamefont {Gibson}}]{Birrell2002}%
  \BibitemOpen
  \bibfield  {author} {\bibinfo {author} {\bibfnamefont {J.}~\bibnamefont
  {Birrell}}, \bibinfo {author} {\bibfnamefont {J.~A.}\ \bibnamefont
  {Carlisle}}, \bibinfo {author} {\bibfnamefont {O.}~\bibnamefont {Auciello}},
  \bibinfo {author} {\bibfnamefont {D.~M.}\ \bibnamefont {Gruen}},\ and\
  \bibinfo {author} {\bibfnamefont {J.~M.}\ \bibnamefont {Gibson}},\ }\bibfield
   {title} {\bibinfo {title} {{Morphology and electronic structure in
  nitrogen-doped ultrananocrystalline diamond}},\ }\href@noop {} {\bibfield
  {journal} {\bibinfo  {journal} {Applied Physics Letters}\ }\textbf {\bibinfo
  {volume} {81}},\ \bibinfo {pages} {2235} (\bibinfo {year}
  {2002})}\BibitemShut {NoStop}%
\bibitem [{\citenamefont {Bhattacharyya}\ \emph {et~al.}(2001)\citenamefont
  {Bhattacharyya}, \citenamefont {Auciello}, \citenamefont {Birrell},
  \citenamefont {Carlisle}, \citenamefont {Curtiss}, \citenamefont {Goyette},
  \citenamefont {Gruen}, \citenamefont {Krauss}, \citenamefont {Schlueter},
  \citenamefont {Sumant},\ and\ \citenamefont {Zapol}}]{Bhattacharyya2001}%
  \BibitemOpen
  \bibfield  {author} {\bibinfo {author} {\bibfnamefont {S.}~\bibnamefont
  {Bhattacharyya}}, \bibinfo {author} {\bibfnamefont {O.}~\bibnamefont
  {Auciello}}, \bibinfo {author} {\bibfnamefont {J.}~\bibnamefont {Birrell}},
  \bibinfo {author} {\bibfnamefont {J.~A.}\ \bibnamefont {Carlisle}}, \bibinfo
  {author} {\bibfnamefont {L.~A.}\ \bibnamefont {Curtiss}}, \bibinfo {author}
  {\bibfnamefont {A.~N.}\ \bibnamefont {Goyette}}, \bibinfo {author}
  {\bibfnamefont {D.~M.}\ \bibnamefont {Gruen}}, \bibinfo {author}
  {\bibfnamefont {A.~R.}\ \bibnamefont {Krauss}}, \bibinfo {author}
  {\bibfnamefont {J.}~\bibnamefont {Schlueter}}, \bibinfo {author}
  {\bibfnamefont {A.}~\bibnamefont {Sumant}},\ and\ \bibinfo {author}
  {\bibfnamefont {P.}~\bibnamefont {Zapol}},\ }\bibfield  {title} {\bibinfo
  {title} {{Synthesis and characterization of highly-conducting nitrogen-doped
  ultrananocrystalline diamond films}},\ }\href
  {https://doi.org/10.1063/1.1400761} {\bibfield  {journal} {\bibinfo
  {journal} {Applied Physics Letters}\ }\textbf {\bibinfo {volume} {79}},\
  \bibinfo {pages} {1441} (\bibinfo {year} {2001})}\BibitemShut {NoStop}%
\bibitem [{\citenamefont {Beloborodov}\ \emph {et~al.}(2006)\citenamefont
  {Beloborodov}, \citenamefont {Zapol}, \citenamefont {Gruen},\ and\
  \citenamefont {Curtiss}}]{Zapol2006}%
  \BibitemOpen
  \bibfield  {author} {\bibinfo {author} {\bibfnamefont {I.~S.}\ \bibnamefont
  {Beloborodov}}, \bibinfo {author} {\bibfnamefont {P.}~\bibnamefont {Zapol}},
  \bibinfo {author} {\bibfnamefont {D.~M.}\ \bibnamefont {Gruen}},\ and\
  \bibinfo {author} {\bibfnamefont {L.~A.}\ \bibnamefont {Curtiss}},\
  }\bibfield  {title} {\bibinfo {title} {Transport properties of n-type
  ultrananocrystalline diamond films},\ }\href@noop {} {\bibfield  {journal}
  {\bibinfo  {journal} {Phys. Rev. B}\ }\textbf {\bibinfo {volume} {74}},\
  \bibinfo {pages} {235434} (\bibinfo {year} {2006})}\BibitemShut {NoStop}%
\bibitem [{\citenamefont {Achatz}\ \emph {et~al.}(2006)\citenamefont {Achatz},
  \citenamefont {Williams}, \citenamefont {Bruno}, \citenamefont {Gruen},
  \citenamefont {Garrido},\ and\ \citenamefont {Stutzmann}}]{Achatz2006}%
  \BibitemOpen
  \bibfield  {author} {\bibinfo {author} {\bibfnamefont {P.}~\bibnamefont
  {Achatz}}, \bibinfo {author} {\bibfnamefont {O.~A.}\ \bibnamefont
  {Williams}}, \bibinfo {author} {\bibfnamefont {P.}~\bibnamefont {Bruno}},
  \bibinfo {author} {\bibfnamefont {D.~M.}\ \bibnamefont {Gruen}}, \bibinfo
  {author} {\bibfnamefont {J.~A.}\ \bibnamefont {Garrido}},\ and\ \bibinfo
  {author} {\bibfnamefont {M.}~\bibnamefont {Stutzmann}},\ }\bibfield  {title}
  {\bibinfo {title} {{Effect of nitrogen on the electronic properties of
  ultrananocrystalline diamond thin films grown on quartz and diamond
  substrates}},\ }\href@noop {} {\bibfield  {journal} {\bibinfo  {journal}
  {Physical Review B}\ }\textbf {\bibinfo {volume} {74}} (\bibinfo {year}
  {2006})}\BibitemShut {NoStop}%
\bibitem [{\citenamefont {Birrell}\ \emph {et~al.}(2003)\citenamefont
  {Birrell}, \citenamefont {Gerbi}, \citenamefont {Auciello}, \citenamefont
  {Gibson}, \citenamefont {Gruen},\ and\ \citenamefont
  {Carlisle}}]{Birrell2003}%
  \BibitemOpen
  \bibfield  {author} {\bibinfo {author} {\bibfnamefont {J.}~\bibnamefont
  {Birrell}}, \bibinfo {author} {\bibfnamefont {J.~E.}\ \bibnamefont {Gerbi}},
  \bibinfo {author} {\bibfnamefont {O.}~\bibnamefont {Auciello}}, \bibinfo
  {author} {\bibfnamefont {J.~M.}\ \bibnamefont {Gibson}}, \bibinfo {author}
  {\bibfnamefont {D.~M.}\ \bibnamefont {Gruen}},\ and\ \bibinfo {author}
  {\bibfnamefont {J.~A.}\ \bibnamefont {Carlisle}},\ }\bibfield  {title}
  {\bibinfo {title} {{Bonding structure in nitrogen doped ultrananocrystalline
  diamond}},\ }\href {https://doi.org/10.1063/1.1564880} {\bibfield  {journal}
  {\bibinfo  {journal} {Journal of Applied Physics}\ }\textbf {\bibinfo
  {volume} {93}},\ \bibinfo {pages} {5606} (\bibinfo {year}
  {2003})}\BibitemShut {NoStop}%
\bibitem [{\citenamefont {Ikeda}\ \emph {et~al.}(2008)\citenamefont {Ikeda},
  \citenamefont {Teii}, \citenamefont {Casiraghi}, \citenamefont {Robertson},\
  and\ \citenamefont {Ferrari}}]{Ikeda2008}%
  \BibitemOpen
  \bibfield  {author} {\bibinfo {author} {\bibfnamefont {T.}~\bibnamefont
  {Ikeda}}, \bibinfo {author} {\bibfnamefont {K.}~\bibnamefont {Teii}},
  \bibinfo {author} {\bibfnamefont {C.}~\bibnamefont {Casiraghi}}, \bibinfo
  {author} {\bibfnamefont {J.}~\bibnamefont {Robertson}},\ and\ \bibinfo
  {author} {\bibfnamefont {A.~C.}\ \bibnamefont {Ferrari}},\ }\bibfield
  {title} {\bibinfo {title} {{Effect of the sp$^\text{2}$ carbon phase on n
  -type conduction in nanodiamond films}},\ }\href@noop {} {\bibfield
  {journal} {\bibinfo  {journal} {Journal of Applied Physics}\ }\textbf
  {\bibinfo {volume} {104}} (\bibinfo {year} {2008})}\BibitemShut {NoStop}%
\bibitem [{\citenamefont {Alcantar-Peña}\ \emph {et~al.}(2016)\citenamefont
  {Alcantar-Peña}, \citenamefont {Montes}, \citenamefont {Arellano-Jimenez},
  \citenamefont {Aguilar}, \citenamefont {Berman-Mendoza}, \citenamefont
  {García}, \citenamefont {Yacaman},\ and\ \citenamefont
  {Auciello}}]{Dallas2016}%
  \BibitemOpen
  \bibfield  {author} {\bibinfo {author} {\bibfnamefont {J.}~\bibnamefont
  {Alcantar-Peña}}, \bibinfo {author} {\bibfnamefont {J.}~\bibnamefont
  {Montes}}, \bibinfo {author} {\bibfnamefont {M.}~\bibnamefont
  {Arellano-Jimenez}}, \bibinfo {author} {\bibfnamefont {J.~O.}\ \bibnamefont
  {Aguilar}}, \bibinfo {author} {\bibfnamefont {D.}~\bibnamefont
  {Berman-Mendoza}}, \bibinfo {author} {\bibfnamefont {R.}~\bibnamefont
  {García}}, \bibinfo {author} {\bibfnamefont {M.}~\bibnamefont {Yacaman}},\
  and\ \bibinfo {author} {\bibfnamefont {O.}~\bibnamefont {Auciello}},\
  }\bibfield  {title} {\bibinfo {title} {Low temperature hot filament chemical
  vapor deposition of ultrananocrystalline diamond films with tunable sheet
  resistance for electronic power devices},\ }\href@noop {} {\bibfield
  {journal} {\bibinfo  {journal} {Diamond and Related Materials}\ }\textbf
  {\bibinfo {volume} {69}},\ \bibinfo {pages} {207 } (\bibinfo {year}
  {2016})}\BibitemShut {NoStop}%
\bibitem [{\citenamefont {Asmussen}\ and\ \citenamefont
  {Reinhard}(2010)}]{DiamondFilmsHandbook}%
  \BibitemOpen
  \bibfield  {author} {\bibinfo {author} {\bibfnamefont {J.}~\bibnamefont
  {Asmussen}}\ and\ \bibinfo {author} {\bibfnamefont {D.~K.}\ \bibnamefont
  {Reinhard}},\ }\href@noop {} {\emph {\bibinfo {title} {Diamond Films
  Handbook}}}\ (\bibinfo {year} {2010})\BibitemShut {NoStop}%
\bibitem [{\citenamefont {Williams}\ \emph {et~al.}(2004)\citenamefont
  {Williams}, \citenamefont {Curat}, \citenamefont {Gerbi}, \citenamefont
  {Gruen},\ and\ \citenamefont {Jackman}}]{Williams2004}%
  \BibitemOpen
  \bibfield  {author} {\bibinfo {author} {\bibfnamefont {O.~A.}\ \bibnamefont
  {Williams}}, \bibinfo {author} {\bibfnamefont {S.}~\bibnamefont {Curat}},
  \bibinfo {author} {\bibfnamefont {J.~E.}\ \bibnamefont {Gerbi}}, \bibinfo
  {author} {\bibfnamefont {D.~M.}\ \bibnamefont {Gruen}},\ and\ \bibinfo
  {author} {\bibfnamefont {R.~B.}\ \bibnamefont {Jackman}},\ }\bibfield
  {title} {\bibinfo {title} {n-type conductivity in ultrananocrystalline
  diamond films},\ }\href@noop {} {\bibfield  {journal} {\bibinfo  {journal}
  {Applied Physics Letters}\ }\textbf {\bibinfo {volume} {85}},\ \bibinfo
  {pages} {1680} (\bibinfo {year} {2004})}\BibitemShut {NoStop}%
\bibitem [{\citenamefont {Pimenta}\ \emph {et~al.}(2007)\citenamefont
  {Pimenta}, \citenamefont {Dresselhaus}, \citenamefont {Dresselhaus},
  \citenamefont {Can{\c{c}}ado}, \citenamefont {Jorio},\ and\ \citenamefont
  {Saito}}]{StudyingDisorder}%
  \BibitemOpen
  \bibfield  {author} {\bibinfo {author} {\bibfnamefont {M.~A.}\ \bibnamefont
  {Pimenta}}, \bibinfo {author} {\bibfnamefont {G.}~\bibnamefont
  {Dresselhaus}}, \bibinfo {author} {\bibfnamefont {M.~S.}\ \bibnamefont
  {Dresselhaus}}, \bibinfo {author} {\bibfnamefont {L.~G.}\ \bibnamefont
  {Can{\c{c}}ado}}, \bibinfo {author} {\bibfnamefont {A.}~\bibnamefont
  {Jorio}},\ and\ \bibinfo {author} {\bibfnamefont {R.}~\bibnamefont {Saito}},\
  }\bibfield  {title} {\bibinfo {title} {{Studying disorder in graphite-based
  systems by Raman spectroscopy}},\ }\href {https://doi.org/10.1039/b613962k}
  {\bibfield  {journal} {\bibinfo  {journal} {Physical Chemistry Chemical
  Physics}\ }\textbf {\bibinfo {volume} {9}},\ \bibinfo {pages} {1276}
  (\bibinfo {year} {2007})}\BibitemShut {NoStop}%
\bibitem [{\citenamefont {Ferrari}\ and\ \citenamefont
  {Robertson}(2000)}]{Ferrari_Robertson2000}%
  \BibitemOpen
  \bibfield  {author} {\bibinfo {author} {\bibfnamefont {A.~C.}\ \bibnamefont
  {Ferrari}}\ and\ \bibinfo {author} {\bibfnamefont {J.}~\bibnamefont
  {Robertson}},\ }\bibfield  {title} {\bibinfo {title} {Interpretation of raman
  spectra of disordered and amorphous carbon},\ }\href@noop {} {\bibfield
  {journal} {\bibinfo  {journal} {Phys. Rev. B}\ }\textbf {\bibinfo {volume}
  {61}},\ \bibinfo {pages} {14095} (\bibinfo {year} {2000})}\BibitemShut
  {NoStop}%
\bibitem [{\citenamefont {Gröning}\ \emph {et~al.}(1999)\citenamefont
  {Gröning}, \citenamefont {Küttel}, \citenamefont {Gröning},\ and\
  \citenamefont
  {Schlapbach}}]{Field_emission_properties_of_nanocrystalline_chemically_vapor_deposited_diamond_films}%
  \BibitemOpen
  \bibfield  {author} {\bibinfo {author} {\bibfnamefont {O.}~\bibnamefont
  {Gröning}}, \bibinfo {author} {\bibfnamefont {O.~M.}\ \bibnamefont
  {Küttel}}, \bibinfo {author} {\bibfnamefont {P.}~\bibnamefont {Gröning}},\
  and\ \bibinfo {author} {\bibfnamefont {L.}~\bibnamefont {Schlapbach}},\
  }\bibfield  {title} {\bibinfo {title} {Field emission properties of
  nanocrystalline chemically vapor deposited-diamond films},\ }\href@noop {}
  {\bibfield  {journal} {\bibinfo  {journal} {Journal of Vacuum Science \&
  Technology B}\ }\textbf {\bibinfo {volume} {17}},\ \bibinfo {pages} {1970}
  (\bibinfo {year} {1999})}\BibitemShut {NoStop}%
\bibitem [{\citenamefont {Pfeiffer}\ \emph {et~al.}(2003)\citenamefont
  {Pfeiffer}, \citenamefont {Kuzmany}, \citenamefont {Salk},\ and\
  \citenamefont {Günther}}]{Pfeiffer2003}%
  \BibitemOpen
  \bibfield  {author} {\bibinfo {author} {\bibfnamefont {R.}~\bibnamefont
  {Pfeiffer}}, \bibinfo {author} {\bibfnamefont {H.}~\bibnamefont {Kuzmany}},
  \bibinfo {author} {\bibfnamefont {N.}~\bibnamefont {Salk}},\ and\ \bibinfo
  {author} {\bibfnamefont {B.}~\bibnamefont {Günther}},\ }\bibfield  {title}
  {\bibinfo {title} {Evidence for trans-polyacetylene in nanocrystalline
  diamond films from h–d isotropic substitution experiments},\ }\href@noop {}
  {\bibfield  {journal} {\bibinfo  {journal} {Applied Physics Letters}\
  }\textbf {\bibinfo {volume} {82}},\ \bibinfo {pages} {4149} (\bibinfo {year}
  {2003})}\BibitemShut {NoStop}%
\bibitem [{\citenamefont {H.~Dong}\ and\ \citenamefont
  {Qian}(2015)}]{CNEnthalpy2015}%
  \BibitemOpen
  \bibfield  {author} {\bibinfo {author} {\bibfnamefont {Q.~Z.}\ \bibnamefont
  {H.~Dong}, \bibfnamefont {A.~R.~Oganov}}\ and\ \bibinfo {author}
  {\bibfnamefont {G.~R.}\ \bibnamefont {Qian}},\ }\bibfield  {title} {\bibinfo
  {title} {The phase diagram and hardness of carbon nitrides},\ }\href@noop {}
  {\bibfield  {journal} {\bibinfo  {journal} {Scientific Reports}\ }\textbf
  {\bibinfo {volume} {5}} (\bibinfo {year} {2015})}\BibitemShut {NoStop}%
\bibitem [{\citenamefont {Gruen}(1999)}]{Gruen_1999}%
  \BibitemOpen
  \bibfield  {author} {\bibinfo {author} {\bibfnamefont {D.~M.}\ \bibnamefont
  {Gruen}},\ }\bibfield  {title} {\bibinfo {title} {Nanocrystalline diamond
  films},\ }\href@noop {} {\bibfield  {journal} {\bibinfo  {journal} {Annual
  Review of Materials Science}\ }\textbf {\bibinfo {volume} {29}},\ \bibinfo
  {pages} {211} (\bibinfo {year} {1999})}\BibitemShut {NoStop}%
\bibitem [{\citenamefont {Baturin}\ \emph {et~al.}(2019)\citenamefont
  {Baturin}, \citenamefont {Nikhar},\ and\ \citenamefont
  {Baryshev}}]{Our_Paper}%
  \BibitemOpen
  \bibfield  {author} {\bibinfo {author} {\bibfnamefont {S.~S.}\ \bibnamefont
  {Baturin}}, \bibinfo {author} {\bibfnamefont {T.}~\bibnamefont {Nikhar}},\
  and\ \bibinfo {author} {\bibfnamefont {S.~V.}\ \bibnamefont {Baryshev}},\
  }\bibfield  {title} {\bibinfo {title} {Field electron emission induced glow
  discharge in a nanodiamond vacuum diode},\ }\href@noop {} {\bibfield
  {journal} {\bibinfo  {journal} {Journal of Physics D}\ }\textbf {\bibinfo
  {volume} {52}},\ \bibinfo {pages} {325301} (\bibinfo {year}
  {2019})}\BibitemShut {NoStop}%
\bibitem [{\citenamefont {Nesl\'adek}\ \emph {et~al.}(1996)\citenamefont
  {Nesl\'adek}, \citenamefont {Meykens}, \citenamefont {Stals}, \citenamefont
  {Van\ifmmode \check{e}\else \v{e}\fi{}\ifmmode~\check{c}\else \v{c}\fi{}ek},\
  and\ \citenamefont {Rosa}}]{Nesladek1996}%
  \BibitemOpen
  \bibfield  {author} {\bibinfo {author} {\bibfnamefont {M.}~\bibnamefont
  {Nesl\'adek}}, \bibinfo {author} {\bibfnamefont {K.}~\bibnamefont {Meykens}},
  \bibinfo {author} {\bibfnamefont {L.~M.}\ \bibnamefont {Stals}}, \bibinfo
  {author} {\bibfnamefont {M.}~\bibnamefont {Van\ifmmode \check{e}\else
  \v{e}\fi{}\ifmmode~\check{c}\else \v{c}\fi{}ek}},\ and\ \bibinfo {author}
  {\bibfnamefont {J.}~\bibnamefont {Rosa}},\ }\bibfield  {title} {\bibinfo
  {title} {Origin of characteristic subgap optical absorption in cvd diamond
  films},\ }\href {https://doi.org/10.1103/PhysRevB.54.5552} {\bibfield
  {journal} {\bibinfo  {journal} {Phys. Rev. B}\ }\textbf {\bibinfo {volume}
  {54}},\ \bibinfo {pages} {5552} (\bibinfo {year} {1996})}\BibitemShut
  {NoStop}%
\bibitem [{\citenamefont {Zapol}\ \emph {et~al.}(2001)\citenamefont {Zapol},
  \citenamefont {Sternberg}, \citenamefont {Curtiss}, \citenamefont
  {Frauenheim},\ and\ \citenamefont {Gruen}}]{Zapol2001}%
  \BibitemOpen
  \bibfield  {author} {\bibinfo {author} {\bibfnamefont {P.}~\bibnamefont
  {Zapol}}, \bibinfo {author} {\bibfnamefont {M.}~\bibnamefont {Sternberg}},
  \bibinfo {author} {\bibfnamefont {L.~A.}\ \bibnamefont {Curtiss}}, \bibinfo
  {author} {\bibfnamefont {T.}~\bibnamefont {Frauenheim}},\ and\ \bibinfo
  {author} {\bibfnamefont {D.~M.}\ \bibnamefont {Gruen}},\ }\bibfield  {title}
  {\bibinfo {title} {Tight-binding molecular-dynamics simulation of impurities
  in ultrananocrystalline diamond grain boundaries},\ }\href@noop {} {\bibfield
   {journal} {\bibinfo  {journal} {Phys. Rev. B}\ }\textbf {\bibinfo {volume}
  {65}},\ \bibinfo {pages} {045403} (\bibinfo {year} {2001})}\BibitemShut
  {NoStop}%
\bibitem [{\citenamefont {Chubenko}\ \emph {et~al.}(2019)\citenamefont
  {Chubenko}, \citenamefont {Baturin},\ and\ \citenamefont
  {Baryshev}}]{Oksana2019}%
  \BibitemOpen
  \bibfield  {author} {\bibinfo {author} {\bibfnamefont {O.}~\bibnamefont
  {Chubenko}}, \bibinfo {author} {\bibfnamefont {S.~S.}\ \bibnamefont
  {Baturin}},\ and\ \bibinfo {author} {\bibfnamefont {S.~V.}\ \bibnamefont
  {Baryshev}},\ }\bibfield  {title} {\bibinfo {title} {Theoretical evaluation
  of electronic density-of-states and transport effects on field emission from
  n-type ultrananocrystalline diamond films},\ }\href@noop {} {\bibfield
  {journal} {\bibinfo  {journal} {Journal of Applied Physics}\ }\textbf
  {\bibinfo {volume} {125}},\ \bibinfo {pages} {205303} (\bibinfo {year}
  {2019})}\BibitemShut {NoStop}%
\end{thebibliography}%

\end{document}